\documentclass[review,authoryear,12pt]{elsarticle}

\usepackage{flushend, cuted}
\usepackage{lipsum}
\usepackage{lingmacros}
\usepackage{tree-dvips}
\usepackage{varioref}
\usepackage{colortbl}
\usepackage{graphicx}
\usepackage{epstopdf}
\usepackage{color,soul}
\usepackage[centertags]{amsmath}
\usepackage{amsfonts}
\usepackage{multicol}
\usepackage{multirow}
\usepackage{hhline}
\usepackage{amssymb}
\usepackage{newlfont}
\usepackage{mathrsfs}
\usepackage{mathtools}
\usepackage{mathrsfs}
\usepackage{cite}
\usepackage{multirow}
\usepackage{nomencl}   
\usepackage{fancyhdr}
\usepackage[caption=false,font=footnotesize]{subfig}
\usepackage{arydshln}
\usepackage{slashbox}
\usepackage{float}
\usepackage{placeins}
\usepackage{tabularx}
\usepackage[table]{xcolor}
\usepackage{amsthm}

\journal{Ultrasound in Medicine and Biology}

\begin{document}

\begin{frontmatter}



\title{Spatio-temporal normalized cross-correlation \\for estimation of the displacement field \\in ultrasound elastography}





\author[Affil1]{Morteza Mirzaei}
\author[Affil1]{Amir Asif}
\author[Affil2]{Maryse Fortin}
\author[Affil1,Affil2]{Hassan Rivaz \corref{cor1}}
\address[Affil1]{Department of Electrical and Computer Engineering, Concordia University, Montreal, Quebec, Canada}
\address[Affil2]{PERFORM Centre, Concordia University, Montreal, Quebec,
	Canada}
\cortext[cor1]{Corresponding Author: Hassan Rivaz, Department of Electrical and Computer Engineering, Concordia University, EV5.235, 1455 Maisonneuve west, Montreal, H3G 1M8; Email, hrivaz@ece.concordia.ca; Phone: 514-848-2424 ext. 8741}

\begin{abstract}
This paper introduces a novel technique to estimate tissue displacement in quasi-static elastography. A major challenge in elastography is estimation of displacement (also referred to time-delay estimation) between pre-compressed and post-compressed ultrasound data. Maximizing normalized cross correlation (NCC) of ultrasound radio-frequency (RF) data of the pre- and post-compressed images is a popular technique for strain estimation due to its simplicity and computational efficiency. Several papers have been published to increase the accuracy and quality of displacement estimation based on NCC. All of these methods use spatial windows to estimate NCC, wherein displacement magnitude is assumed to be constant within each window. In this work, we extend this assumption along the temporal domain to exploit neighboring samples in both spatial and temporal directions.
This is important since traditional and ultrafast ultrasound machines are, respectively, capable of imaging at more than $30$ frame per second (fps) and $1000$ fps. We call our method spatial temporal normalized cross correlation (STNCC)  and show that it substantially outperforms NCC using simulation, phantom and \textit{in-vivo} experiments. 

\end{abstract}


\end{frontmatter}

\pagebreak








\section*{Introduction}
\label{intro}
Ultrasound imaging is one of the most  commonly used imaging modalities since it is inexpensive, safe and convenient. Ultrasound elastography estimates biomechanical properties of the tissue and can substantially improve the capabilities of ultrasound imaging in both diagnosis and image-guided interventions. Elastography methods can reveal different mechanical properties such as viscosity or Poisson's ratio, but imaging elastic properties of the tissue is the most-widely used technique~\citep{5.overview}. Elastography has been used in imaging breast~\citep{1.overview,2.overview,3.overview,breast} and prostate cancer ~\citep{4.overview}  as well as 
investigation of liver health~\citep{41.overview,machineliver} and surgical treatment of liver cancer~\citep{surger,surger2,vargese1,liverlas}.

Estimation of tissue displacement due to an internal or external force is at the heart of all ultrasound elastography methods~\citep{sarvaz}. Elastography methods that are based on internal or endogenous deformation are often based on the pumping action of the heart which generates waves  in the surrounding tissue. Mechanical properties of the cardiac tissue can be measured based on velocity of this wave~\citep{endogen1,endogen2,endogen3}.
In the case of external excitation, there are different techniques for exciting tissue and measuring its mechanical property but they can be broadly grouped into dynamic and quasi-static elastography. Dynamic methods such as shear wave imaging (SWI)  \citep{7.overview, mahmo, lastshear} and acoustic radiation force imaging (ARFI) \citep{8.overview} can provide quantitative mechanical properties of tissue. Both SWI and ARFI use Acoustic Radiation Force (ARF) to generate displacement in the tissue.

Quasi-static elastography often generates the displacement in the tissue by simply pressing the probe against the tissue. The core idea of quasi static approach that is also known as compression elastography is introduced in~\citep{9.overview} but the concept of this technique is not a new one and estimation of tissue hardness by hand palpation is an ancient technique~\citep{6.overview}.
The main reason for name of quasi-static is that the velocity of deformation is very low such that static mechanics can be assumed~\citep{10.overview}. This technique does not require additional hardware other than an ultrasound machine, and as such, is very convenient and has even been applied in image-guided surgery~\citep{3d2} and radiotherapy~\citep{surger2}. Compared to SWI and ARFI, displacements in quasi-static elastography are usually substantially larger, leading to a larger signal to noise ratio in displacement estimation. The disadvantage is that it cannot readily generate quantitative tissue properties and an inverse problem approach should also be applied to infer quantitative properties in tissue~\citep{inverse1,inverse2,quanti}.

This paper entails estimation of tissue displacement, and as such, can be applied to almost all elastography methods. However, we focus on free-hand palpation quasi-static elastography, which involves slowly compressing the tissue with the ultrasound probe. Low cost and availability are two  advantages of free-hand palpation ultrasound elastography~\citep{freehand1,freehand2}. In this method, the movement of the probe is largely in the axial direction and the main goal is to compute strain and deformation in the axial direction. However, even pure axial compression of probe will deform the tissue in all directions. Although axial deformation has most of useful elasticity information, but lateral displacement can also be calculated~\citep{lateral,glue,interpolation2,laterallast}. Estimation of out-of-plane deformation is currently not possible from two dimensional ultrasound images, and custom-made probes~\citep{outofplaneprobe} or three-dimensional ultrasound imaging is needed~\citep{3d2, 3d1, 3d3}. Deformation estimation is most accurate in the axial direction since ultrasound resolution is very high in this direction, and as such, often only axial displacement is estimated in elastography.

Estimation of tissue displacement is often referred to as time delay estimation (TDE), which relies on raw radio-frequency (RF) data. Since one sample of RF data does not provide enough information to calculate displacement, most methods are based on dividing the RF data into several overlapping windows and calculating the displacement of each window~\citep{windowcomparision}. The underlying assumption here is that displacement of all samples within the window is the same, and therefore, additional information from the neighboring samples is exploited to calculate the displacement of the sample at the center of the window. This additional information helps reduce the estimation variance. 

 Maximization of the normalized cross correlation (NCC) of windows was one of the first approaches used for TDE, which is still a very popular approach because it is easy-to-implement and computationally efficient~\citep{ncc2,ncc1,NCCPSO}. Phase correlation wherein zero crossing of phase determines displacement \citep{phase1,phase2} and sum of absolute difference of windows \citep{sumofabsolute} are other major window-based techniques for elastography.
  
  Window-based techniques are easy to implement, but one of the most important disadvantages of these algorithms is false peaks. False peaks occur when a secondary NCC peak or zero crossing of phase or sum of absolute difference, exceeds true ones. False peaks are a common error in window-based elastography methods since all windows of post compressed image should be searched to find the best match. To overcome false peaks, time-domain cross correlation with prior estimates (TDPE) is introduces in \citep{ncc1}. In TPDE, only a small part of post compressed image should be searched for correlated window and the searching area is limited to a neighborhood around the previous time-delay estimate. By utilizing TDPE, the problem of false peaks can be addressed but still window-based algorithms are sensitive to   signal de-correlation, which can be caused by the out of plane or lateral displacement which, is a common problem especially in free-hand palpation. Another major source for signal de-correlation is blood flow and other biological motions that are common in \textit{in-vivo} data.

In all of the aforementioned studies, the RF lines of just two images are compared with each other and the displacement fields across small spatial windows are assumed to be constant. 
Inspired by~\citep{zhang2008spacetime}, we extend this assumption to the temporal domain in this work. We consider the cine ultrasound RF data as three-dimensional, where the third dimension is the time domain. 
 We maximize NCC in between three-dimensional windows, and therefore, we name our proposed algorithm as spatial temporal normalized cross correlation (STNCC). This simple and intuitive idea substantially improves results of TDE. It is important to note that although the windows that we utilize to calculate NCC are three-dimensional, the estimated displacement field is two-dimensional.    

STNCC is more robust to signal de-correlation compared to NCC as shown in the simulation experiments. We also show that as the amplitude of noise increases, STNCC exhibits much less susceptibility as compared to NCC. In addition, STNCC is less sensitive to the window size in comparison to NCC. 


This paper is organized as follows. The STNCC method is presented in the next section. Simulation, phantom and \textit{in-vivo} experiments of back muscle and liver are studied in the Results Section. The results of the STNCC method are compared against traditional NCC. Discussions of the results and avenues for future work are presented in the Discussion Section, and the paper is concluded in the Conclusion Section.

\section*{Methods}
Most  elastography methods consider two images $I_1$ and $I_2$ as pre- and post-compressed images, and calculate displacement of tissue using RF data of these images. The pre-compressed image is divided into several windows, and for each window,  one should look for a window in the post compressed image that maximizes  NCC as it is shown in Figure 1. 

\begin{figure}
	\centering
	\subfloat[]{{\includegraphics[width=3.9cm]{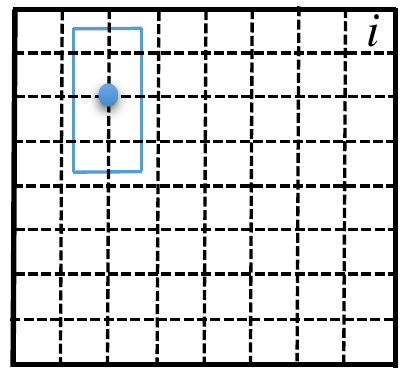} }}
	\qquad
	\subfloat[]{{\includegraphics[width=3.9cm]{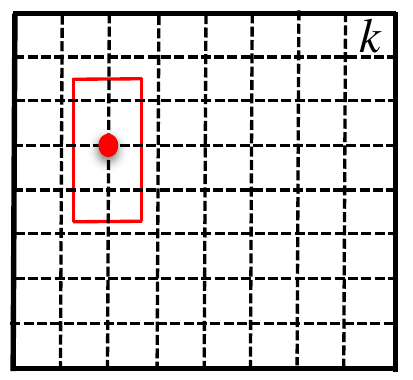} }}
	\caption{Two frames of ultrasound images corresponding to (a) pre- and (b) post-compression.
		Vertical dashed lines represent RF lines and intersection of vertical and
		horizontal lines represent RF samples. The images are severely downsampled for visual illustration; a typical RF frame has many more samples.  To find the displacement of the sample marked with a blue circle, the blue window around that sample is considered for calculating a similarity metric (usually NCC). The red sample indicates the corresponding sample in the post-compression image.}
\end{figure} 
NCC for two windows $A$ and $B$  is calculated as eqn (\ref{eq1}),
\begin{equation}
\dfrac{\varSigma_{j=1}^{j=1+W}A(j)B(j)}{\sqrt{\varSigma_{j=1}^{j=1+W}A(j)^2}\sqrt{\varSigma_{j=1}^{j=1+W}B(j)^2}},
\label{eq1}
\end{equation}
where $W$ is the number of samples in the windows and $j$ represent samples of windows. The peak of NCC corresponds to the displacement of windows in the pre-compressed image. Maximization of NCC only provides an integer displacement estimate, and interpolation should be performed to find a more accurate sub-pixel displacement estimate \citep{interpolation,interpolation2,interpolation3}.

In this paper a novel technique is introduced to use temporal information.
Hence instead of two windows, two three-dimensional boxes should be considered as shown in Figure 2.
\begin{figure}
	\centering
	{{\includegraphics[width=8.5cm]{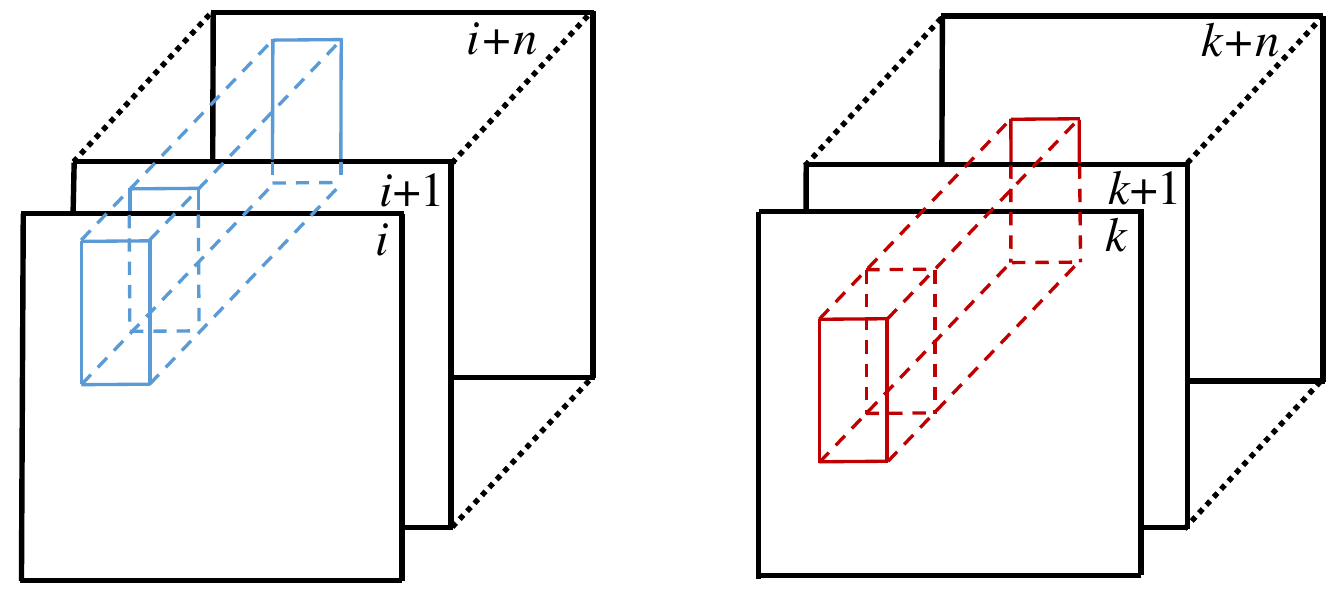} }}
	\caption{Two sequence of images used for spatial and
		temporal estimation of  normalized correlation. The similarity metric (NCC in this work) is computed using the data in the 3D red and blue boxes.}
\end{figure}
 In this technique one should look for a box in the second sequence that has the maximum NCC with the box of first sequence and peak of NCC represents displacement of the center of first box.
The only assumption of this algorithm is that all samples within the box have equal displacements. This is a good assumption since the frame rate of ultrasound machines are more than $30$ fps (more than $1000$ fps if plane-wave imaging is used) and consecutive frames and their displacement will be very close to each other.  By considering $n$ frames for each box, the NCC of the two boxes is defined as eqn (\ref{eq2}),
\begin{equation}
\begin{array}{l}
\dfrac{\varSigma_{l=1}^{l=n}\varSigma_{j=1}^{j=1+W}A_l(j)B_l(j)}{\sqrt{\varSigma_{l=1}^{l=n}\varSigma_{j=1}^{j=1+W}A_l(j)^2}} \\
\times\dfrac{1}{\sqrt{\varSigma_{l=1}^{l=n}\varSigma_{j=1}^{j=1+W}B_l(j)^2}}
\end{array}
\label{eq2}
\end{equation}
where $A_l$ and $B_l$ are windows in the $l^{th}$ frames of first and second boxes. $W$ is the number of samples in a 2D window and $j$ show samples of 2D windows.
The peak of STNCC provides an integer displacement estimate and have to be interpolated to generate a subpixel displacement estimate.  
To avoid false peaks, search area of this algorithm is limited similar to~\citep{ncc1}.
By calculating the displacement field, strain of the tissue can be determined by differentiating displacement field in the axial direction. Differentiating amplifies the noise, and therefore, least square techniques are common method to obtain the strain field. Kalman filter is also used to improve the quality of strain estimation~\citep{dpam}.

\section*{Results}
\label{Results}
In this section, results of the proposed STNCC method are presented and compared against NCC using Filed II~\citep{fielii} and  finite elements method (FEM) simulations, phantom and \textit{in-vivo} data from back muscle and liver. 
Signal to noise ratio (SNR) and contrast to noise ratio (CNR)  are used to provide quantitative means for assessing the proposed method according to eqn (\ref{eq3}),
\begin{equation}
\begin{array}{l}
SNR=\dfrac{\bar{s}}{\sigma},\\
\textrm{CNR}=\sqrt{\dfrac{2(\bar{s}_b-\bar{s}_t)^2}{\sigma_b^2+\sigma_t^2}}, 
\end{array}
\label{eq3}
\end{equation}
where $\bar{s}_t$ and $\bar{s}_b$ are the spatial strain average of the target and
background, $\sigma_b^2$ and $\sigma_t^2$ are the spatial strain variance of the
target and background, and $\bar{s}$ and $\sigma$ are the spatial average and
variance of an arbitrary window in the strain image, respectively.

In all simulations and  experiments, $7$ frames are considered for STNCC and outputs of STNCC are compared with strain of middle frames thats are estimated by NCC.

\subsection*{Simulation Results}
A simulated  phantom is generated by utilizing the Field II  ultrasound simulation software~\citep{fielii}. FEM-based deformations are computed using  the ABAQUS  software package (Providence, RI, USA). The simulated phantom is homogenous except for a cylindrical inclusion with zero stiffness which is placed in the middle of phantom as an inclusion. The inclusion simulates a blood vein that easily compresses under force. 
The phantom is compressed by $0.5\%$, and compression rate between two consecutive frames are $0.02\%$. The ground truth strain is shown in Figure 3 where the white part represents the inclusion.

\begin{figure}
	\centering
	{{\includegraphics[width=5cm]{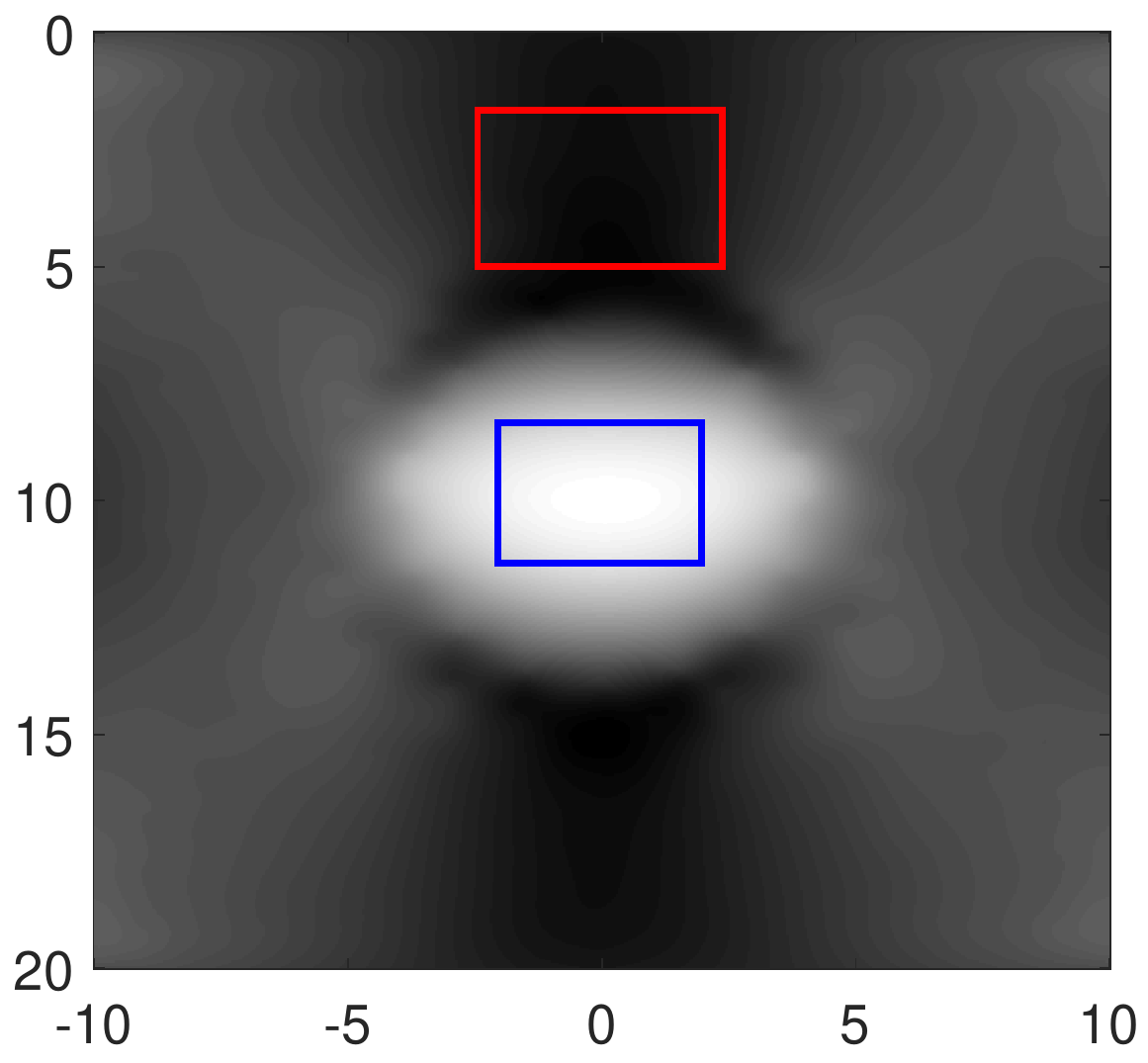} }}
	\caption{Ground truth strain in the simulation phantom. The displacement is estimated using the  ABAQUS FEM software. The red and blue windows are considered respectively as the background and foreground windows for calculation of CNR. The red window is considered to calculate SNR.}
\end{figure}

To make simulation experiment more realistic, images are normalized as eqn (\ref{eq4}),
\begin{equation}
I_{ij}=\dfrac{I_{ij}}{\textrm{max}_{i,j}(I_{ij})}
\label{eq4}
\end{equation}
and uniform noises are added to images in three steps with maximum magnitude of $0.3, 0.5$ and $0.7$. Strains are then calculated by STNCC and NCC with $86\%$ overlap of windows and 3 point parabolic interpolation to find the $2\textrm{D}$ sub-sample location of the correlation peak.    
Figure 4 shows outputs of STNCC and NCC for different levels of noise. 
\begin{figure}
	\centering
	\subfloat[]{{\includegraphics[width=3.9cm]{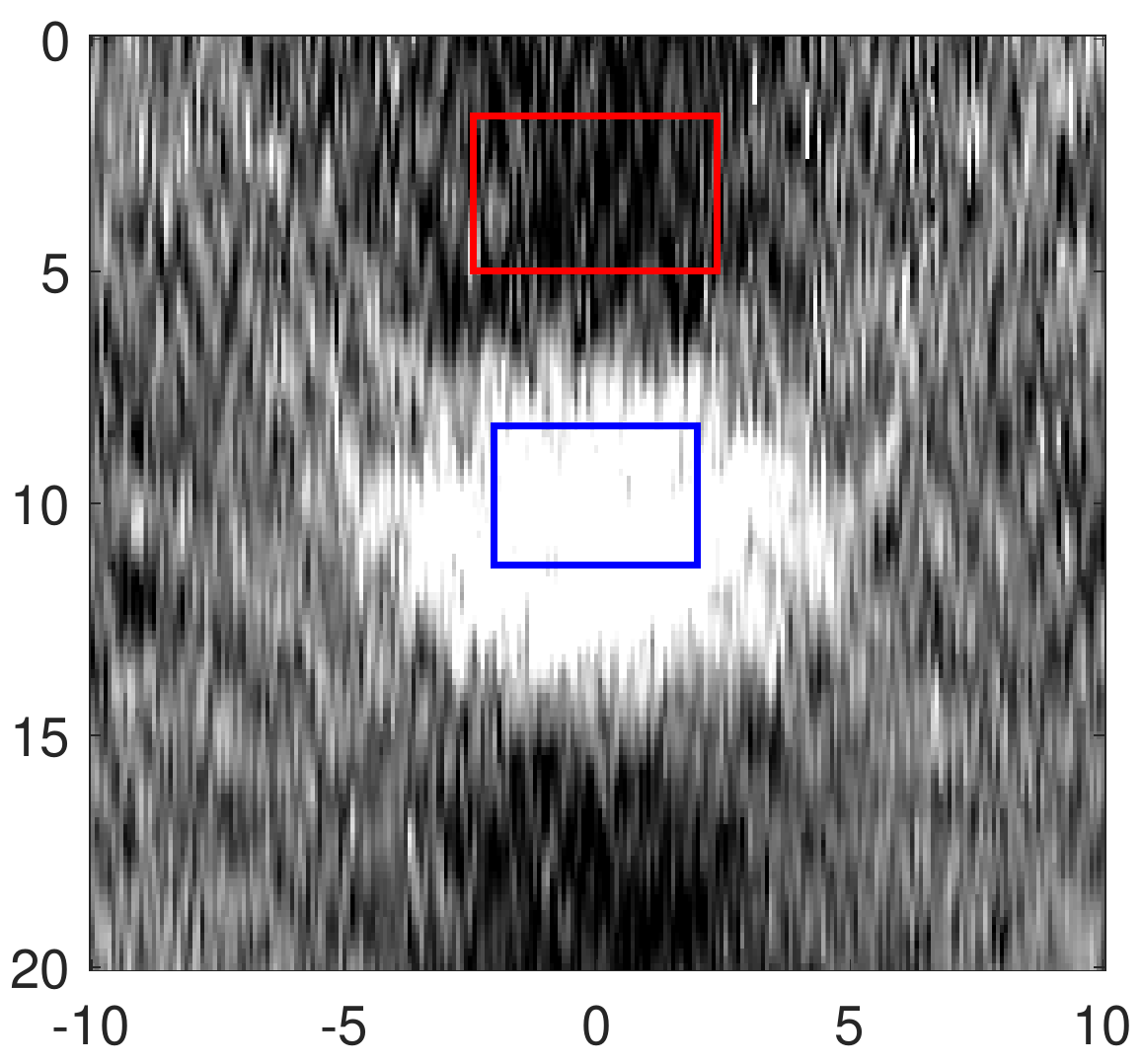} }}	
	\subfloat[]{{\includegraphics[width=3.9cm]{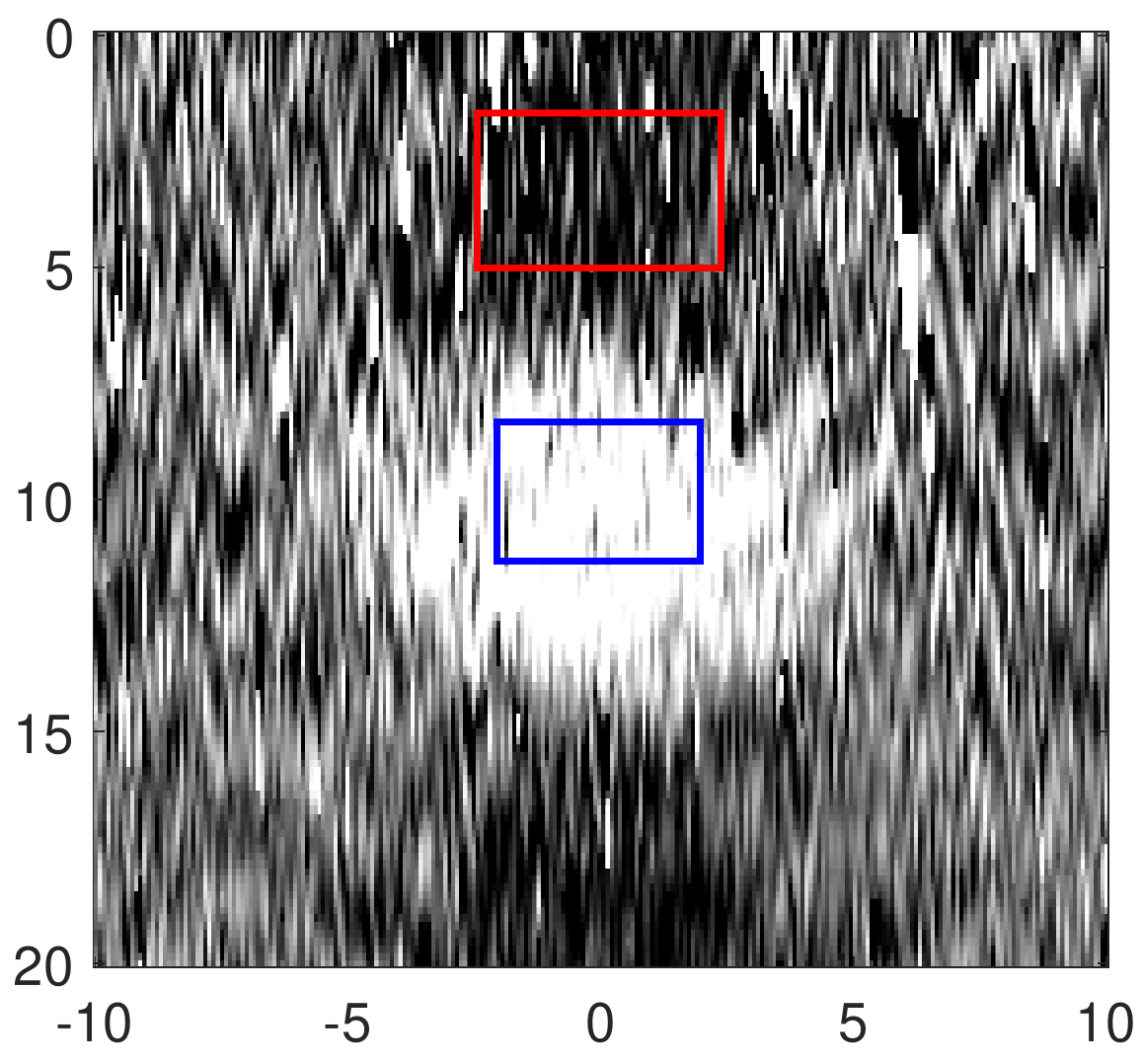} }}%
		\subfloat[]{{\includegraphics[width=3.9cm]{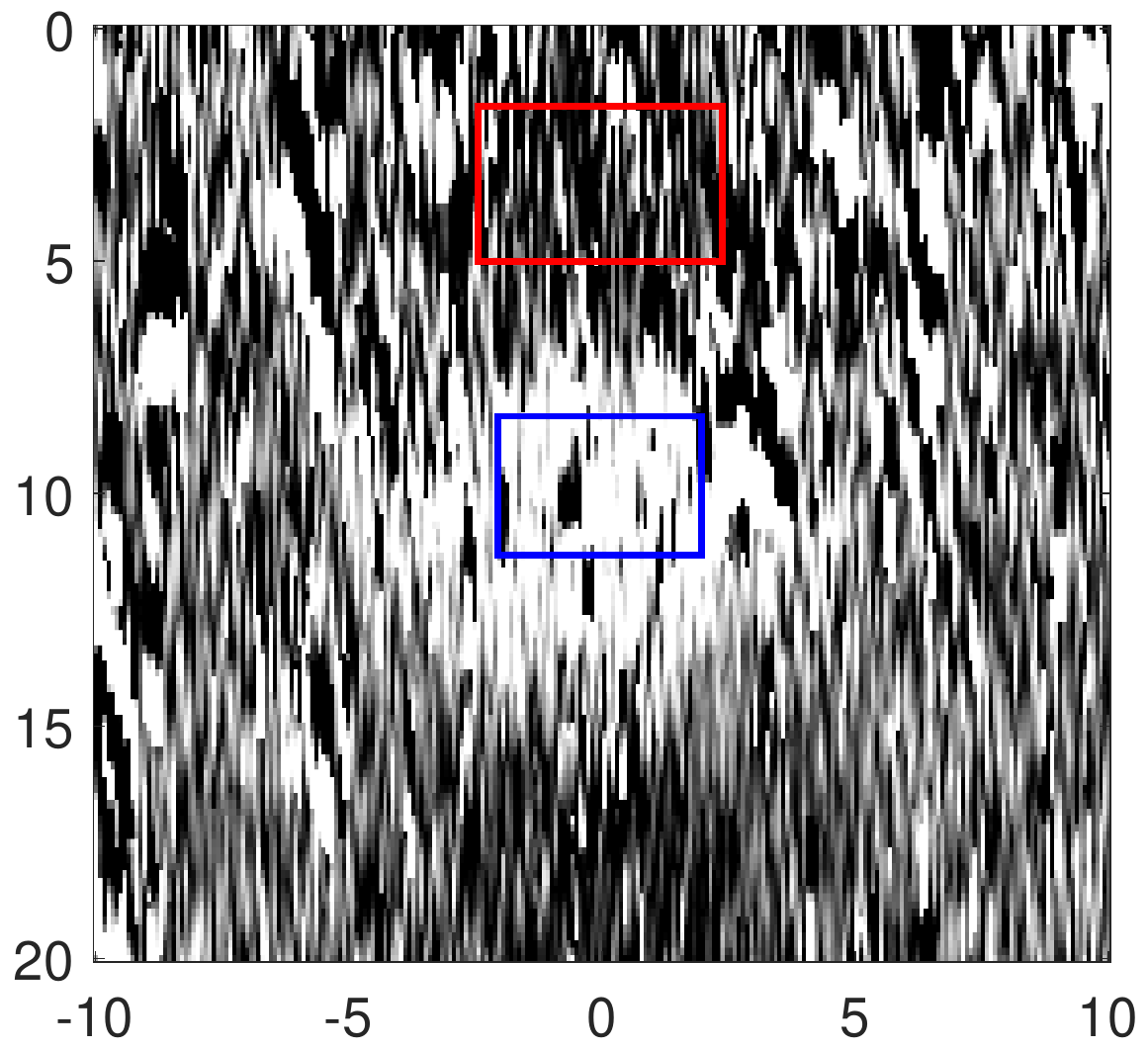} }}
	\qquad
	\subfloat[]{{\includegraphics[width=3.9cm]{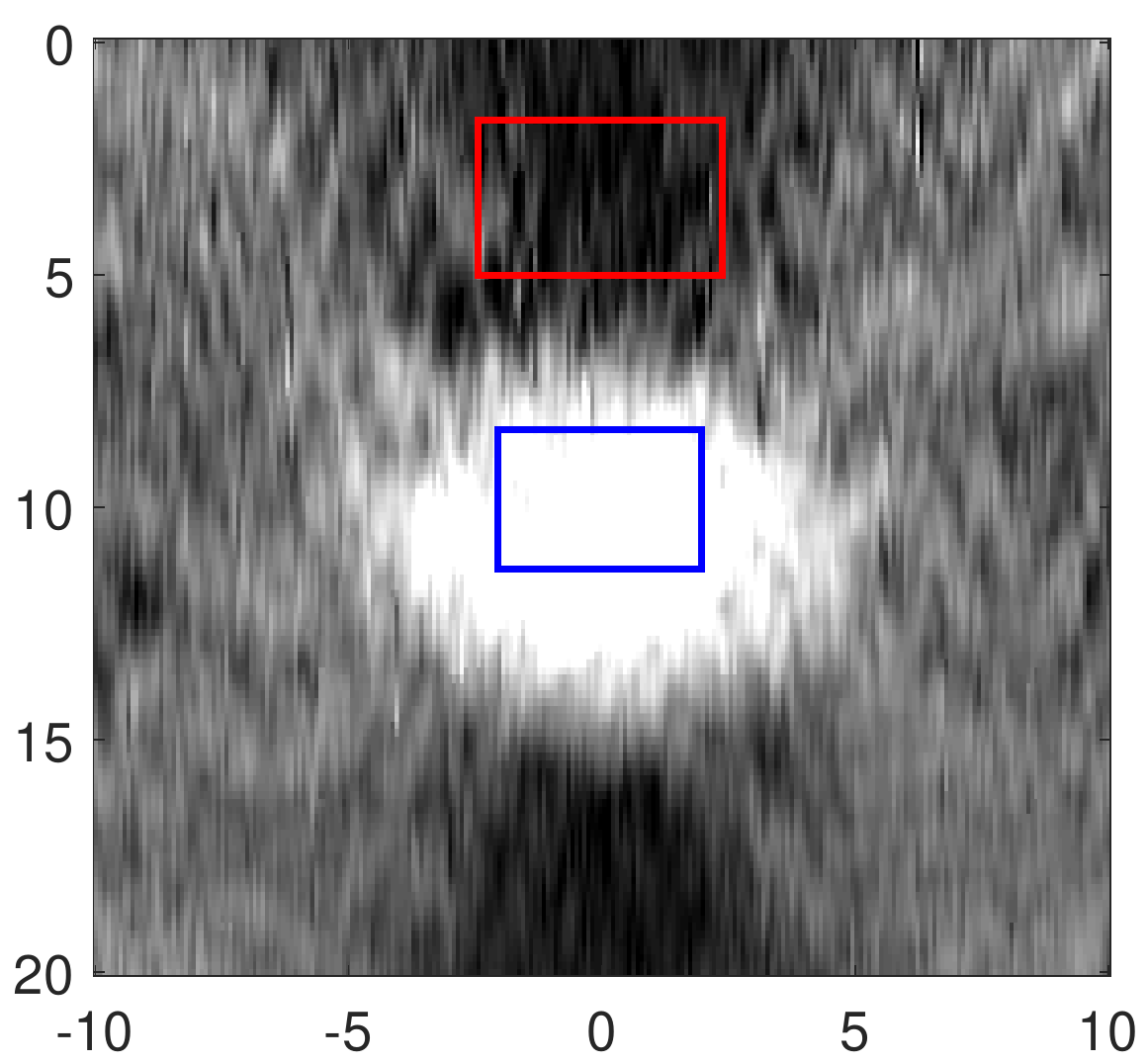} }}
	\subfloat[]{{\includegraphics[width=3.9cm]{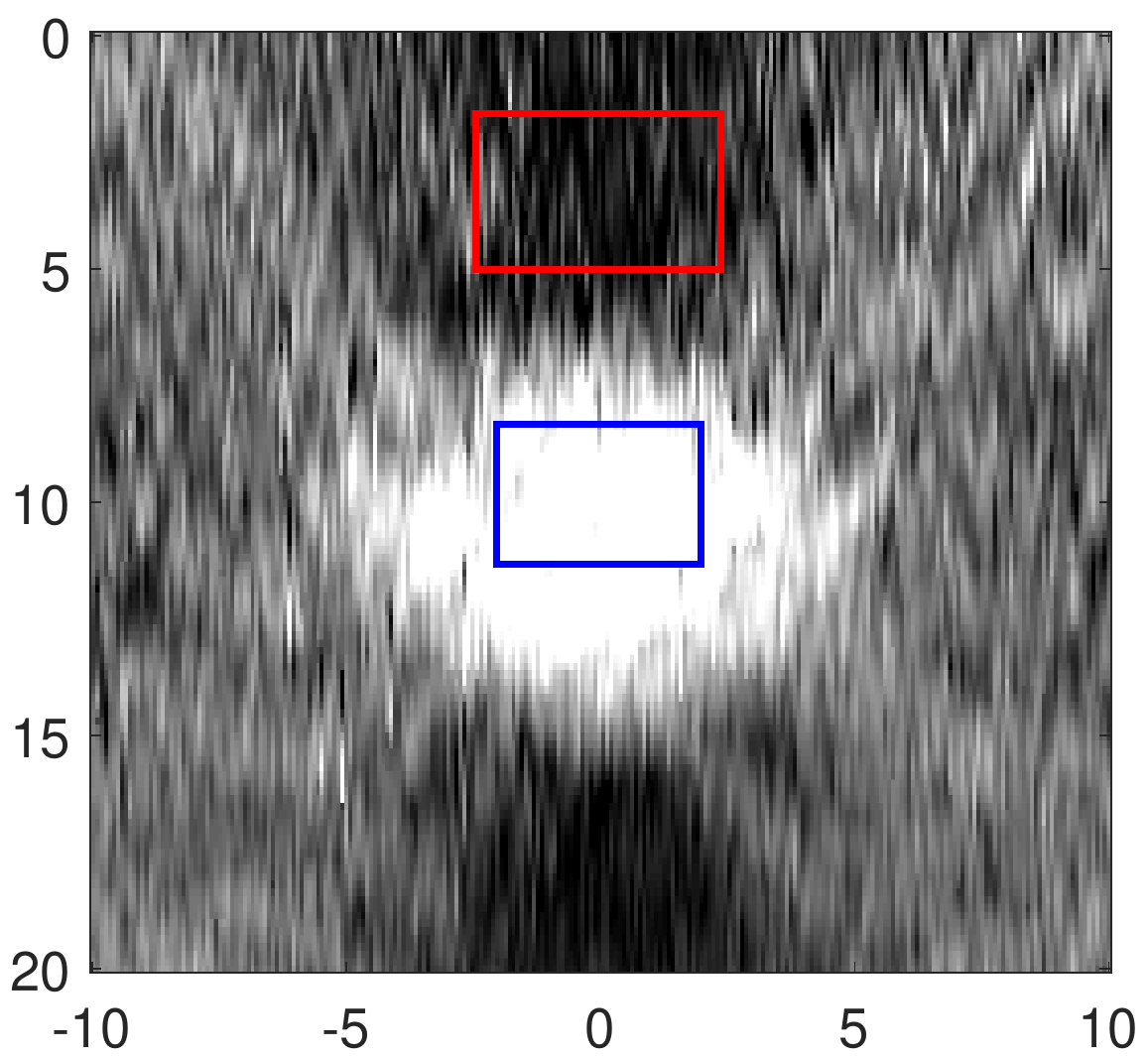} }}
	\subfloat[]{{\includegraphics[width=3.9cm]{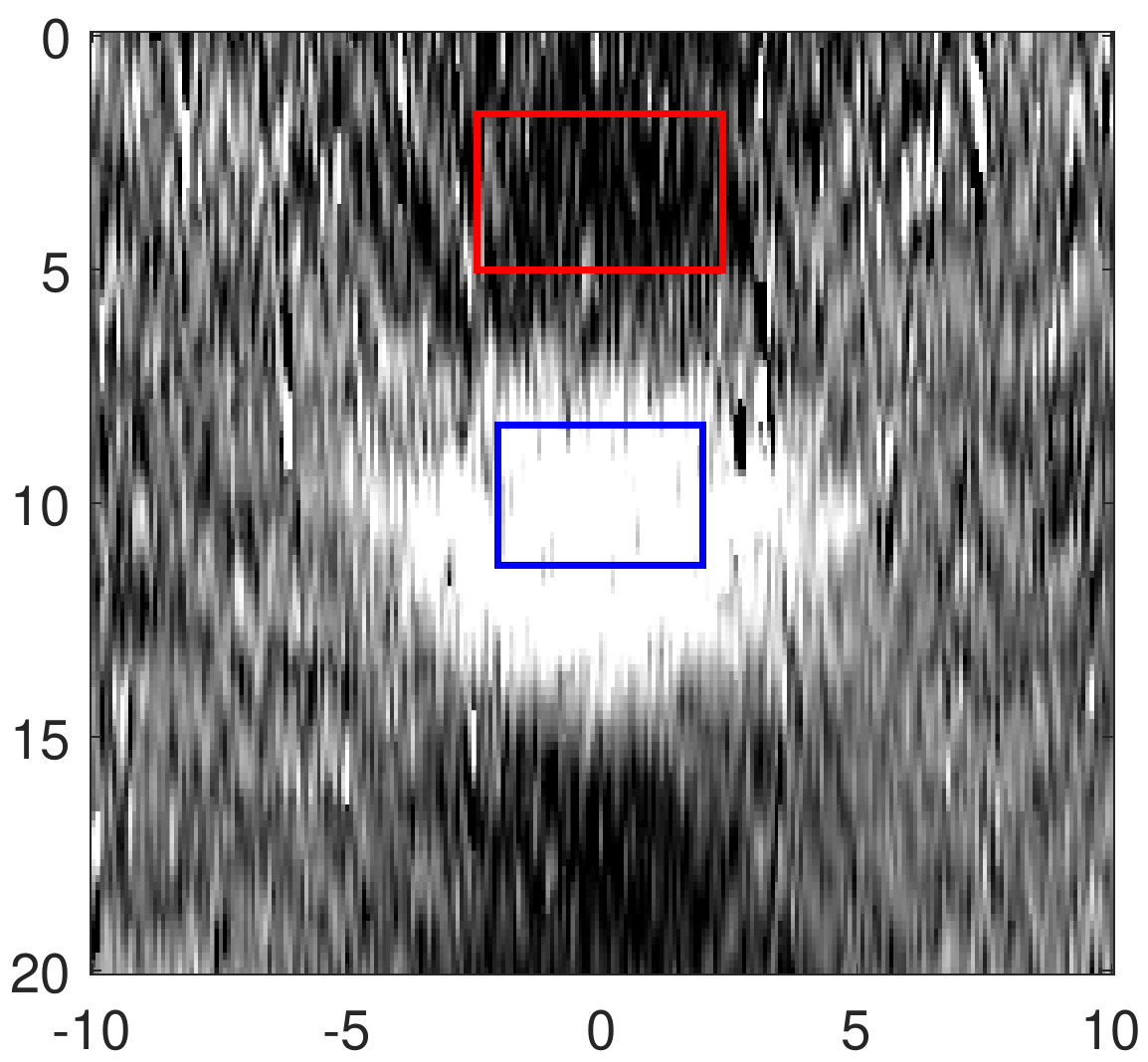} }}
	\qquad
	\subfloat[]{{\includegraphics[width=3.9cm]{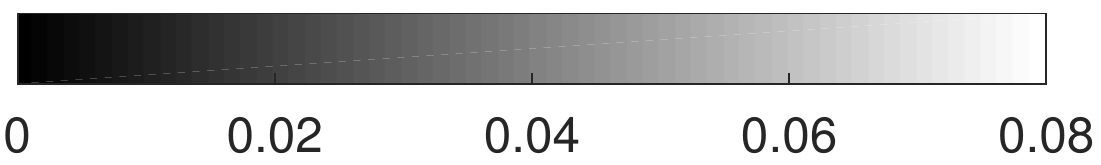} }}
	\caption{Strain images of the simulation phantom calculated using NCC and STNCC.
		The first row shows strain images that are calculated using NCC, and the second row depicts
		strain images  computed using STNCC. In the first, second and third columns, the maximum amplitude of noise values are 0.3, 0.5 and 0.7 respectively.
	}
\end{figure}
It is clear that results of STNCC is closer to ground truth and outperform results of NCC. For calculating signal to noise ratio and contrast to noise ratio that are represented in Table 1, for each level of noise we estimated strain 100 times with different random noise and averaged SNR and CNR of these 100 experiments. As one can see in Figure 4 and Table 2 not even STNCC outperforms NCC for each range of noise, but also has more robust performance for increasing amplitude of noise. 

\begin{table}
	\caption{Averaged SNR and CNR of 100 strain images of the simulated phantom for different methods and noise levels. Windows that are considered for calculating CNR are shown in blue and red lines in Figures 3 and 4. The red window is considered for SNR.}
\begin{tabular}[c]{lllr}
	\hline		
	&	& SNR & CNR \\
	\hline
	\multirow{3}{*}{Noise$=0.3$}	&	STNCC    & 132.50    & 11.59  \\
	&	NCC   & 39.00  &   6.91   \\
	&	\textbf{Improvement}   & $\mathbf{\%239.74}$  &   $\mathbf{\%67.72}$ \\
	\hline
	\multirow{3}{*}{Noise$=0.5$}	&	STNCC    &  58.74   &  8.45 \\
	&	NCC   & 3.04  &  1.89    \\
	&	\textbf{Improvement}   & $\mathbf{\%1832.23}$  &   $\mathbf{\%347.08}$ \\
	\hline
	\multirow{3}{*}{Noise$=0.7$}	&	STNCC    &  15.33   & 4.62  \\
	&	NCC   & fails  &  fails    \\
	&	\textbf{Improvement}   & -  &   - \\
	\hline		
\end{tabular}\\
\end{table}
\begin{table}
	\caption{Effect of increasing noise on SNR and CNR values.} 
	\begin{tabular}[r]{llcc}
		\hline		
		Variation of noise amplitude	&	Method & \%SNR & \%CNR \\
		\hline
		\multirow{2}{*}{From $0.3$ to $0.5$}	&	STNCC    & -55.66    &  -27.09 \\
		&	NCC   & -92.20   &   -72.64   \\
		\hline		
	\end{tabular} \\
\end{table}
In the next experiment, we compressed the simulated phantom by $1\%$, $1.5\%$ and $2\%$ and repeated the experiment for these amount of compression. For representing CNR, simulation is run 100 times for each case and it is shown in Figure 5 that for all three compression rate and for all three different noise levels, STNCC has better performance than NCC. 
\begin{figure*}
	\centering
	\subfloat[]{{\includegraphics[width=3.9cm]{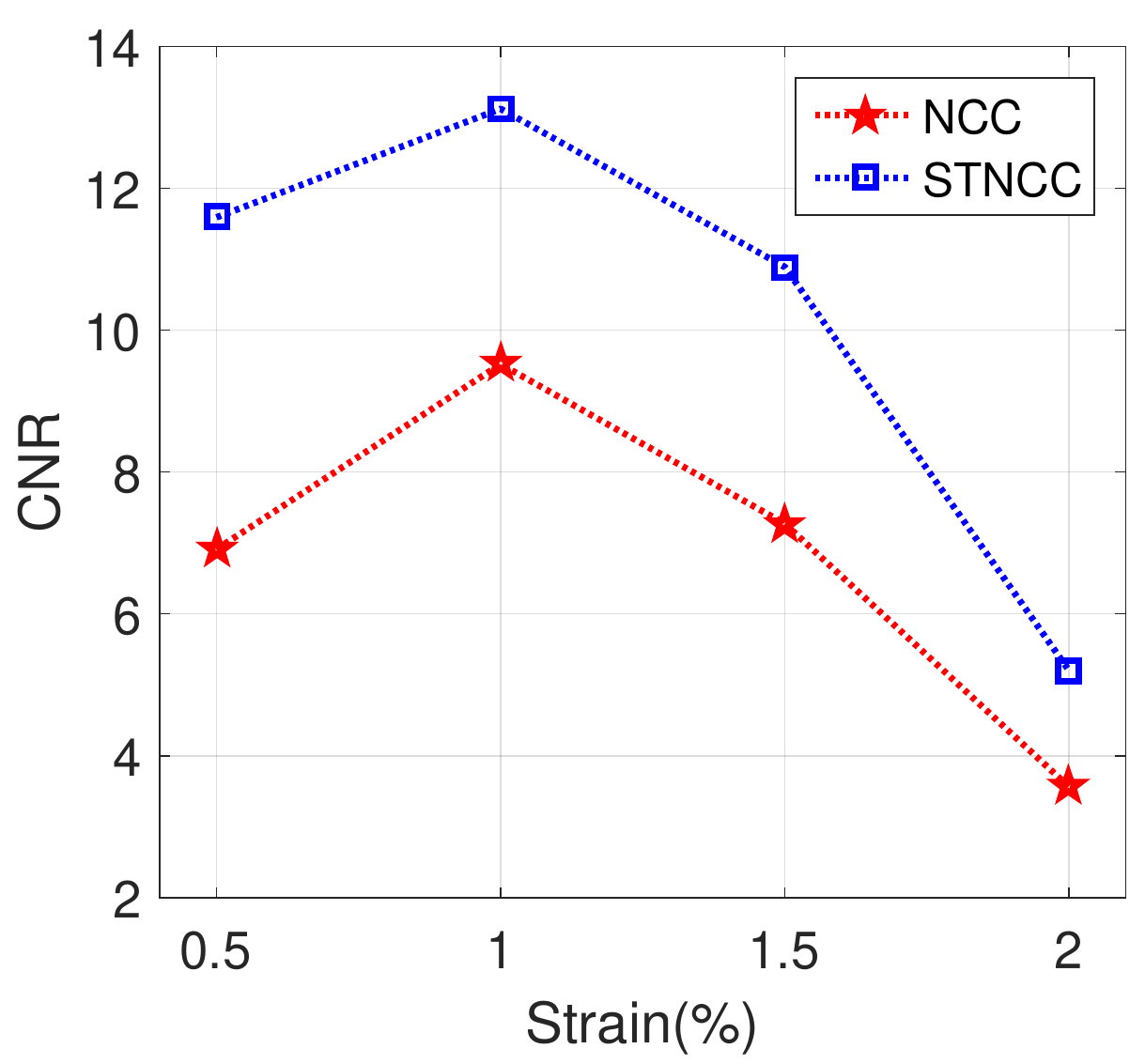} }}%
	\subfloat[]{{\includegraphics[width=3.9cm]{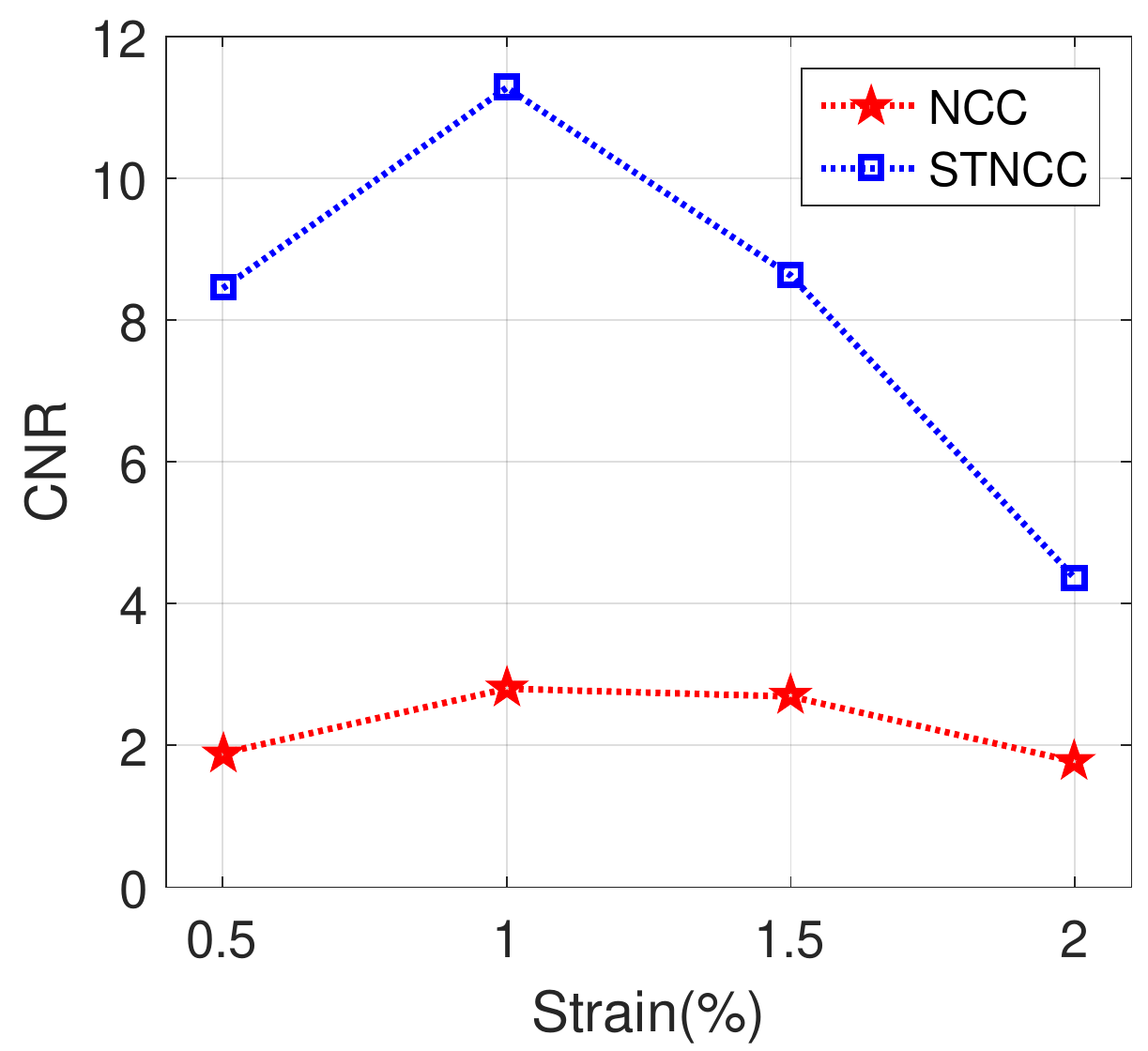} }}
	\subfloat[]{{\includegraphics[width=3.9cm]{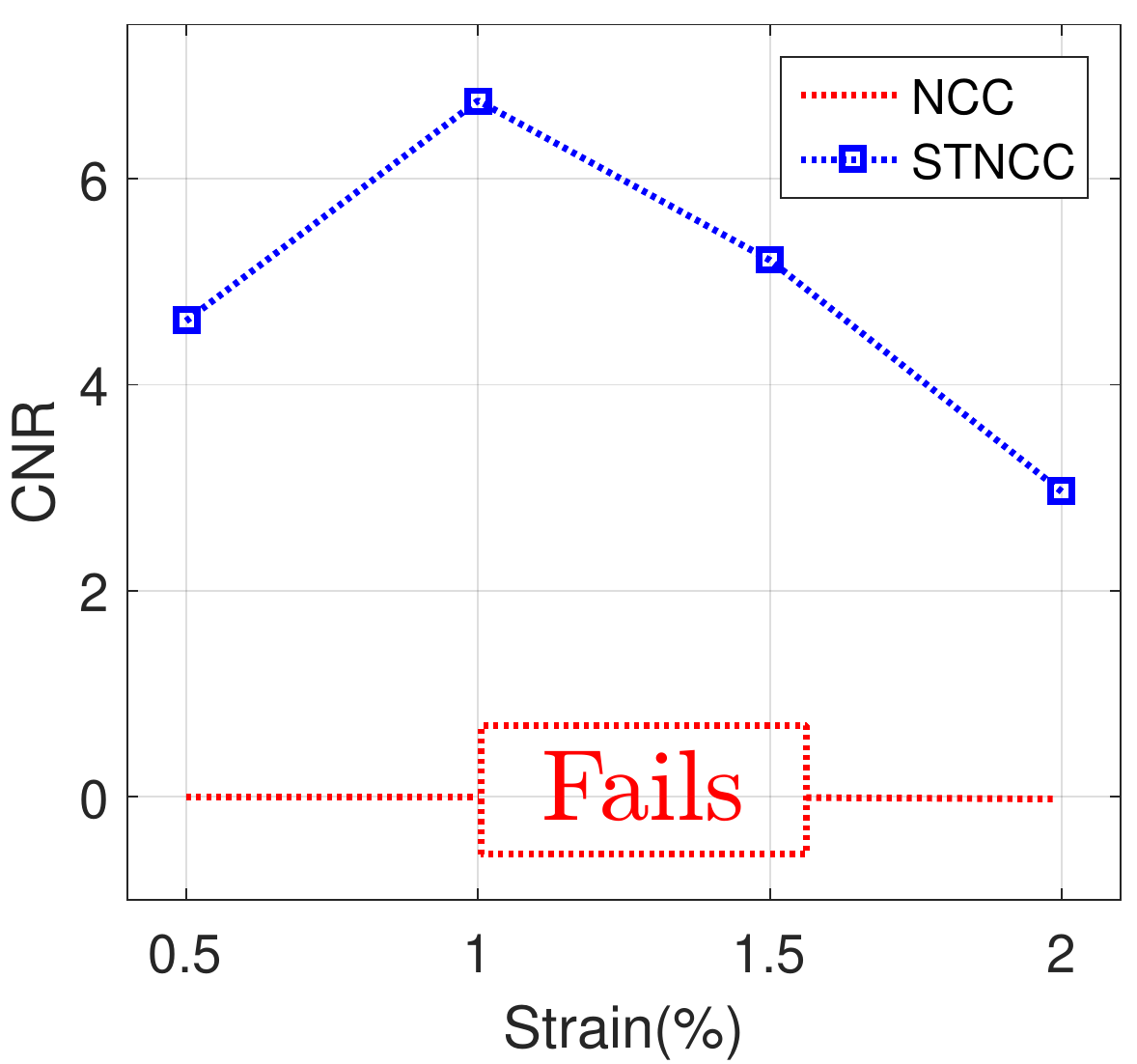} }}
	\caption{CNR values for different levels of compression and noise. The maximum amplitudes of  noise in (a), (b) and (c) are respectively 0.3, 0.5 and 0.7.
	}
\end{figure*}
\subsection*{Phantom Results}

For experimental evaluation, RF data is acquired from an elastography phantom (CIRS tissue simulation \& phantom technology, Norfolk, VA, USA) with an Antares Siemens ultrasound machine (Antaras, Siemens, Issaquah, WA, USA) and VF 13-5 probe at the center frequency of $7.27$ MHz, sampling frequency of $40$ MHz and frame rate of $37$ fps.
Similar to the previous section, the images are normalized and uniform noises are added to images in three steps. Since phantom already includes some noise, the amplitude of added noise is decreased to $0.1, 0.3$ and $0.5$. Strains are calculated by STNCC and NCC and are shown in Figure 6.
\begin{figure}
	\centering
	\subfloat[]{{\includegraphics[width=3.9cm]{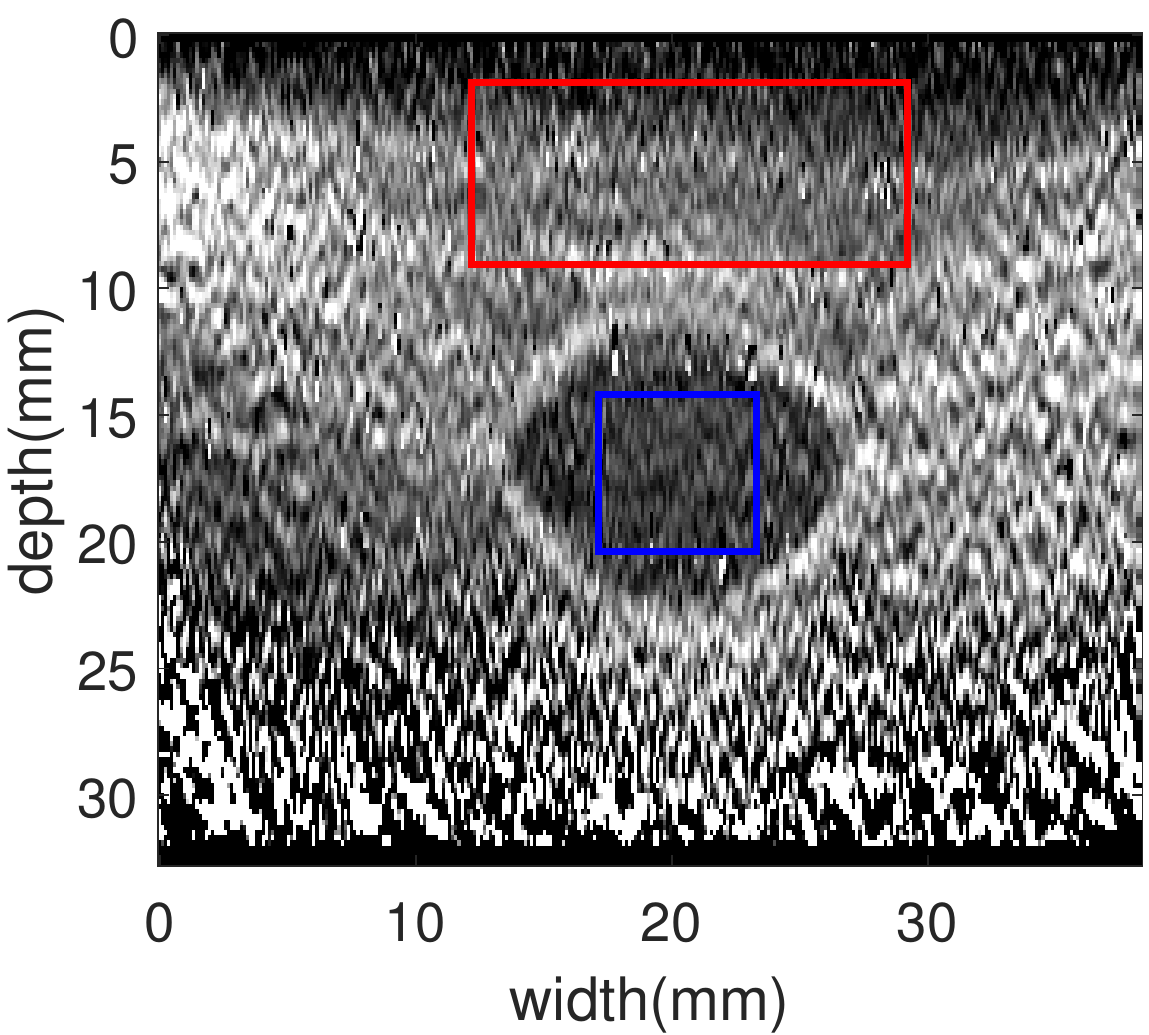} }}%
	\subfloat[]{{\includegraphics[width=3.9cm]{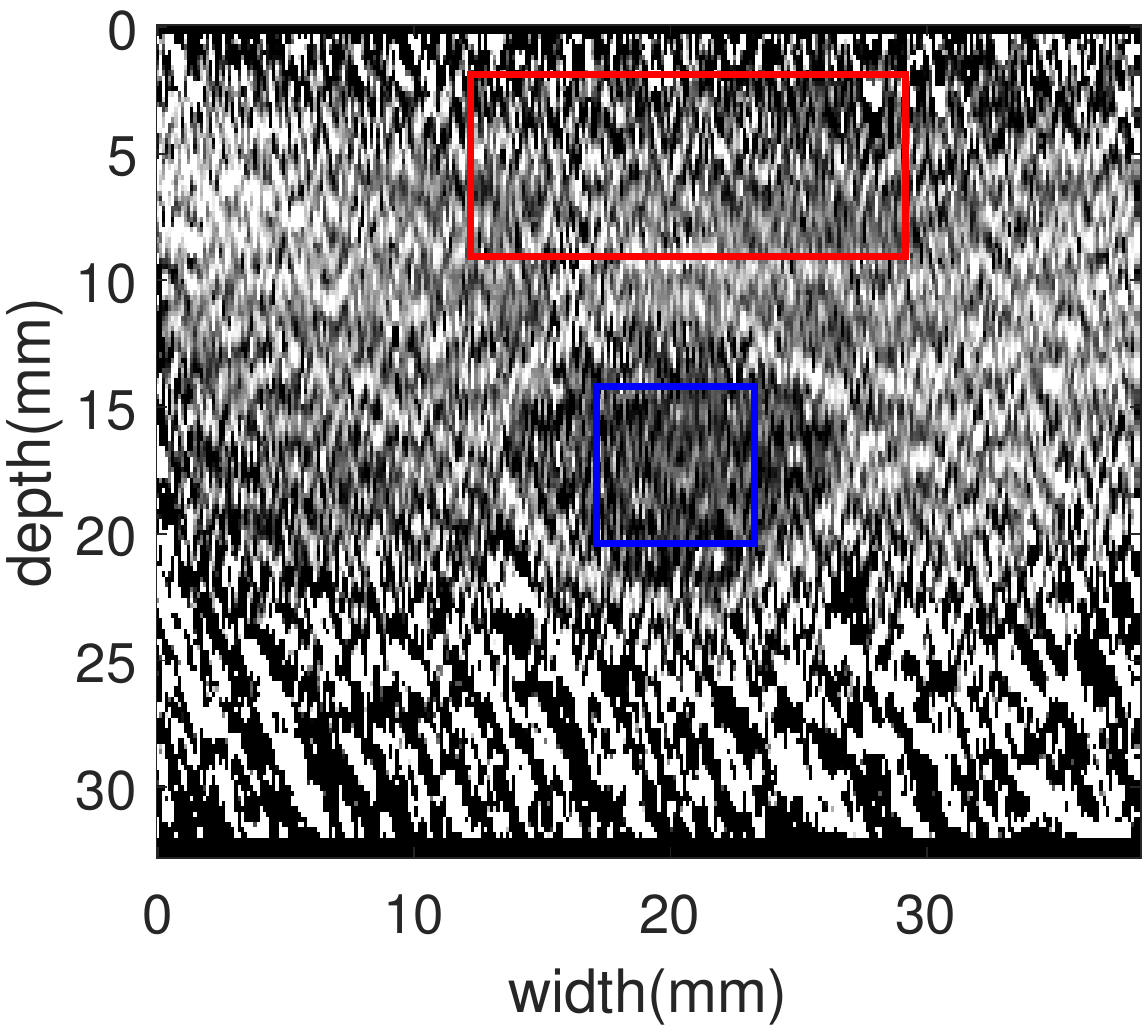} }}%
	\subfloat[]{{\includegraphics[width=3.9cm]{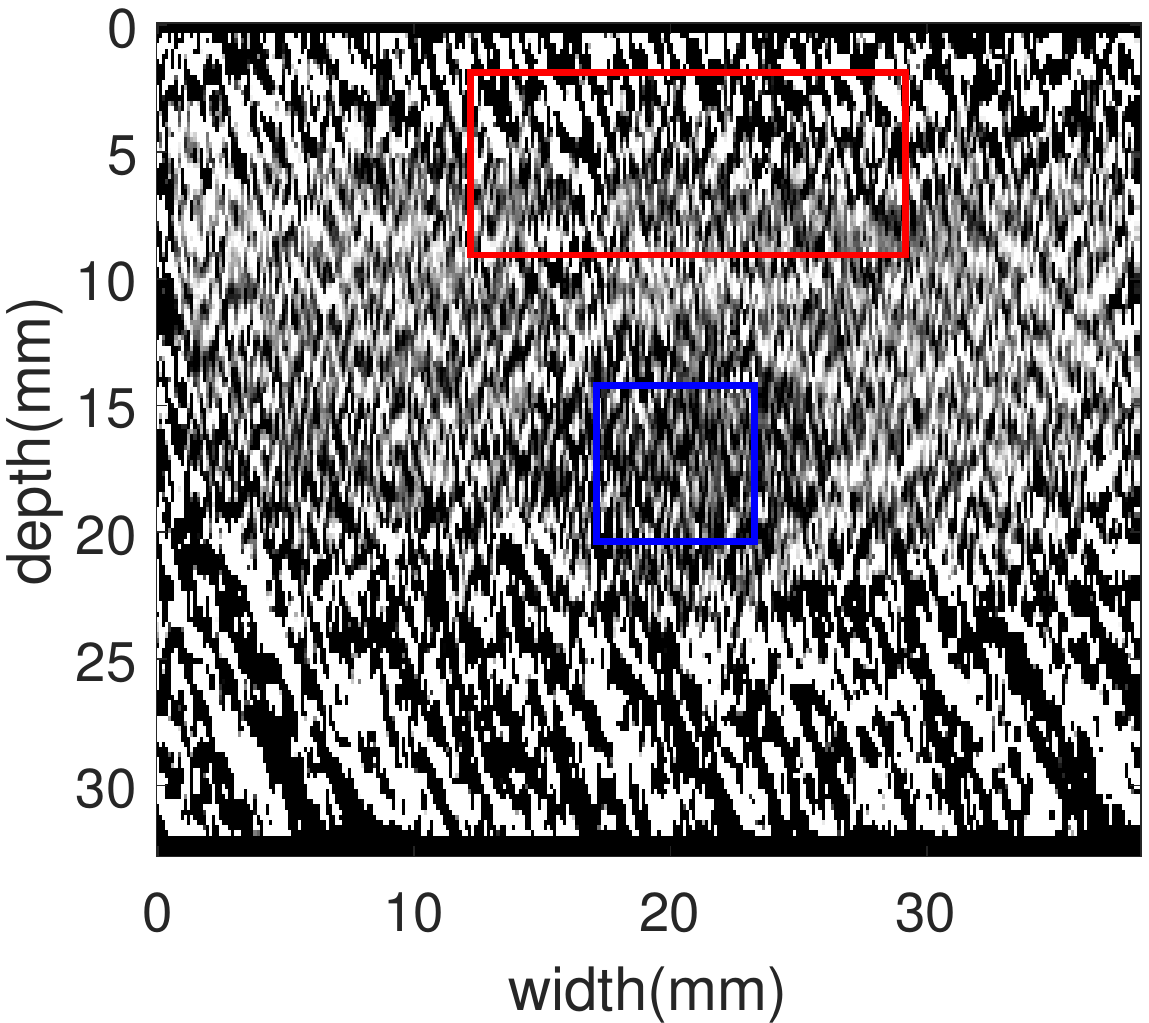} }}
	\qquad
	\subfloat[]{{\includegraphics[width=3.9cm]{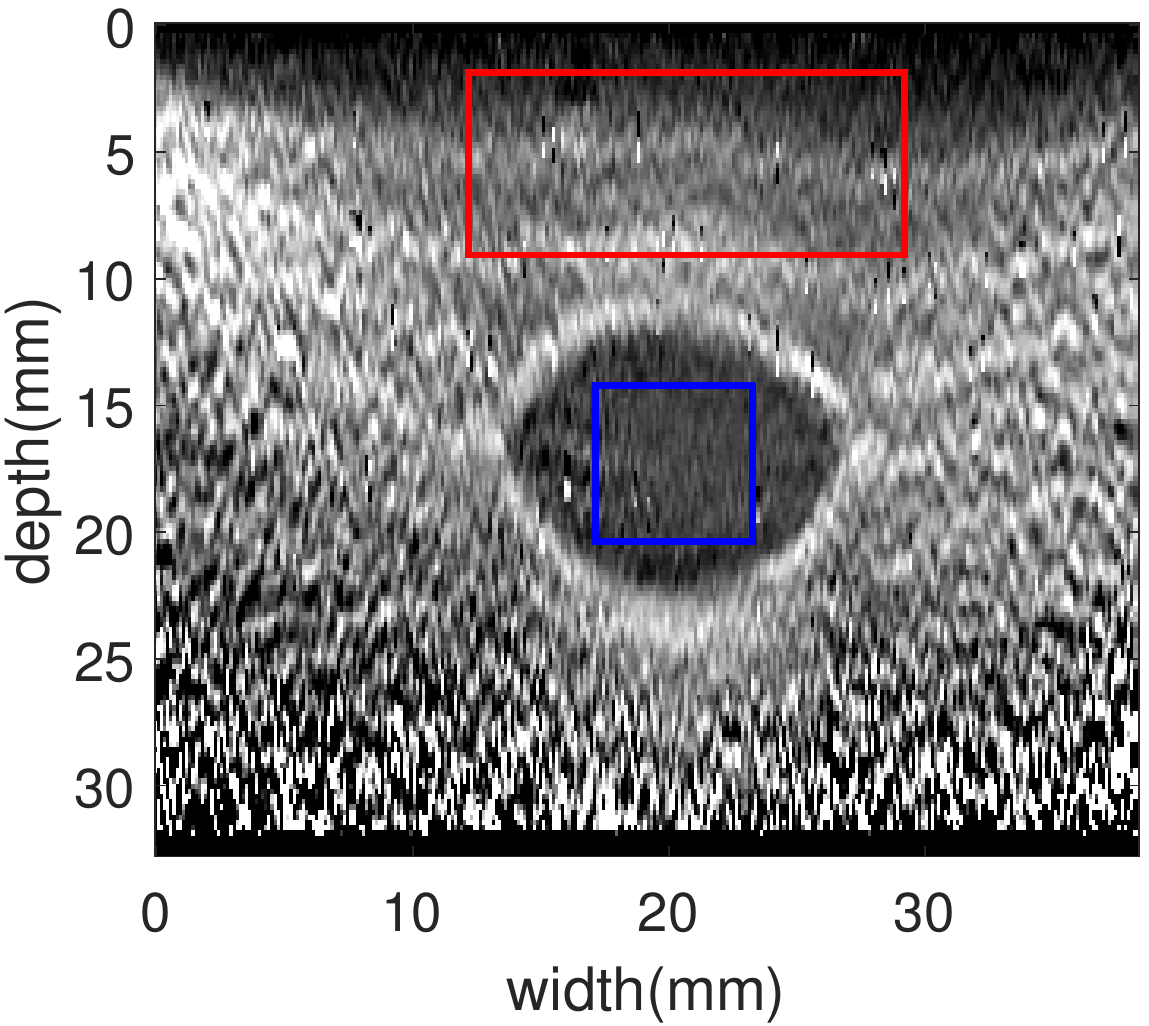} }}
	\subfloat[]{{\includegraphics[width=3.9cm]{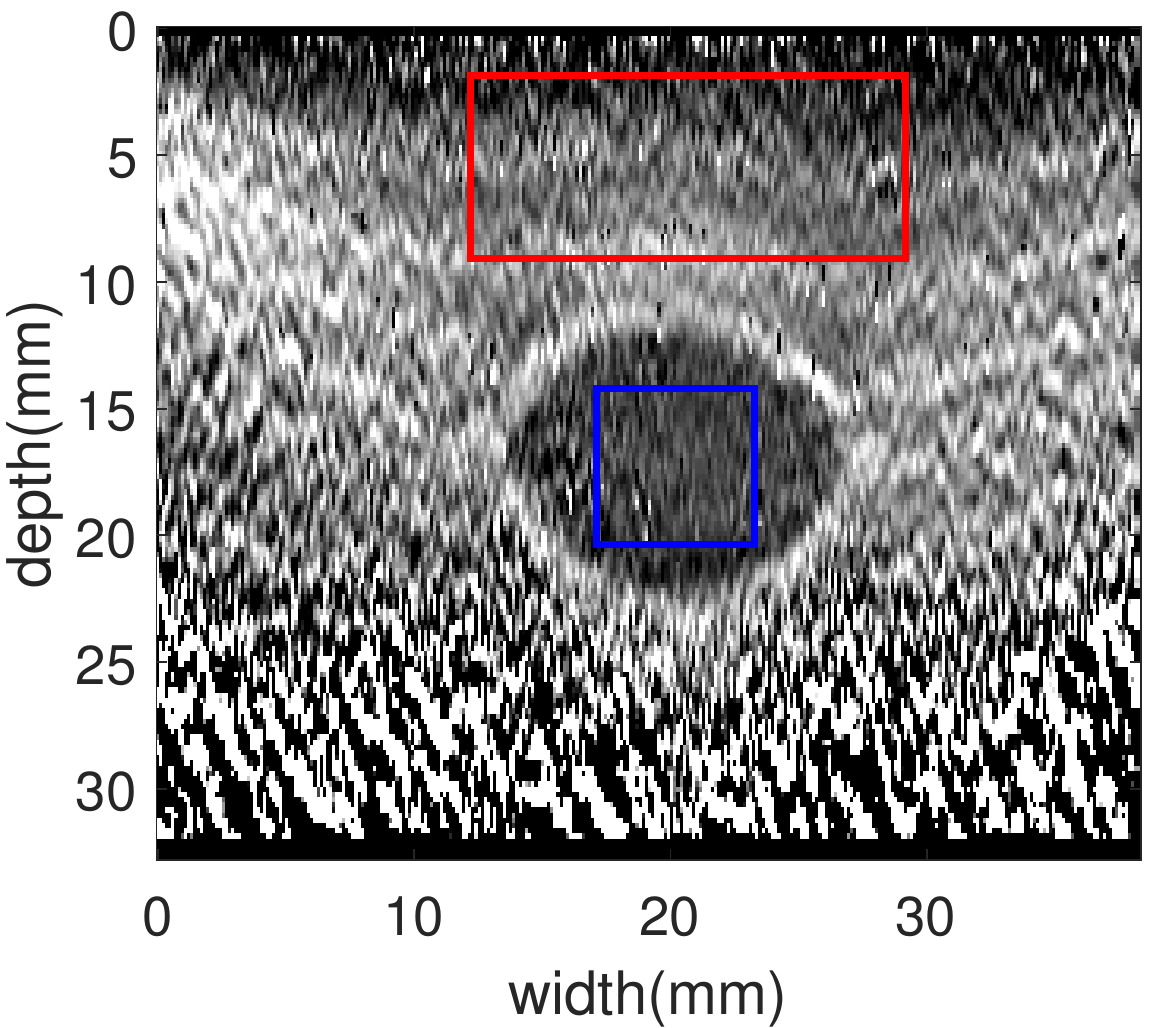} }}
	\subfloat[]{{\includegraphics[width=3.9cm]{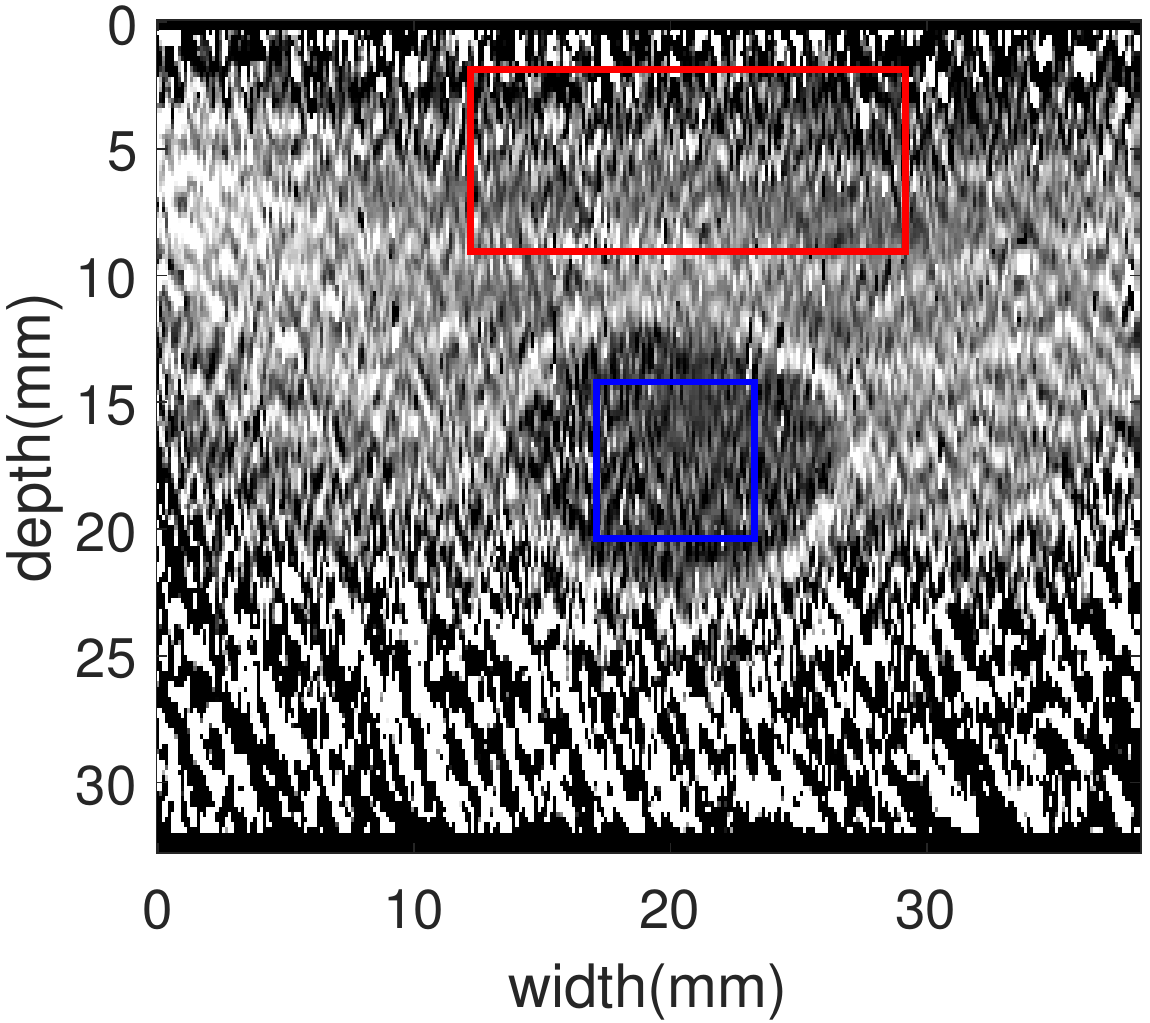} }}
	\qquad
	\subfloat[]{{\includegraphics[width=3.9cm]{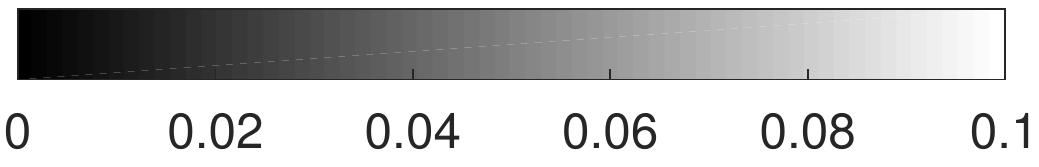} }}
	\caption{Comparison of strains that are calculated using NCC and STNCC for phantom data. The first and second rows show strain images calculated using NCC and STNCC, respectively. In the first, second and third columns, the maximum amplitude of noise values are 0.3, 0.5 and 0.7 respectively.}
\end{figure}

As one can see, results of STNCC outperform NCC and STNCC is more robust to increasing magnitude of noise. For computing SNR and CNR for each level of noise, experiments are repeated for 100 times and averaged SNR and CNR are represented in Table 3. 
\begin{table}
\caption{Average values of SNR and CNR in 100 strain images of the phantom  at different noise levels. Windows that are considered for calculating SNR and CNR are shown in Figure 6  (SNR is computed in the red windows only).}
\begin{tabular}[c]{lllr}
	\hline		
	&	& SNR & CNR \\
	\hline
	\multirow{3}{*}{Noise$=0.1$}	&	STNCC    &  84.29   & 4.18  \\
	&	NCC   & 71.01  &   3.48   \\
	&	\textbf{Improvement}   & $\mathbf{\%18.70}$  &   $\mathbf{\%20.11}$   \\
	\hline
	\multirow{3}{*}{Noise$=0.3$}	&	STNCC    & 47.96   & 3.01  \\
	&	NCC   & 1.03  &  0.48    \\
	&	\textbf{Improvement}  & $\mathbf{\%4556.31}$  &   $\mathbf{\%527.08}$   \\
	\hline
	\multirow{3}{*}{Noise$=0.5$}	&	STNCC    &  1.41   & 0.60  \\
	&	NCC   & fails  &  fails    \\
	&	\textbf{Improvement}   & $-$  &   $-$   \\
	\hline		
\end{tabular}\\ 
\end{table}
Edge spread function of strains obtained by NCC and STNCC are shown in Figure 7. For calculating edge spread function two rectangular with length of 60 and width of 10 pixels are considered in strain of NCC and STNCC as it is shown in Figure 7a-7b. Edge spread function is calculated by averaging intensity of pixels across width of these rectangular and it is clear in Figure 7d-7e that edge spread function of STNCC is smoother than NCC. 
\begin{figure}
	\centering
	\subfloat[]{{\includegraphics[width=3.9cm]{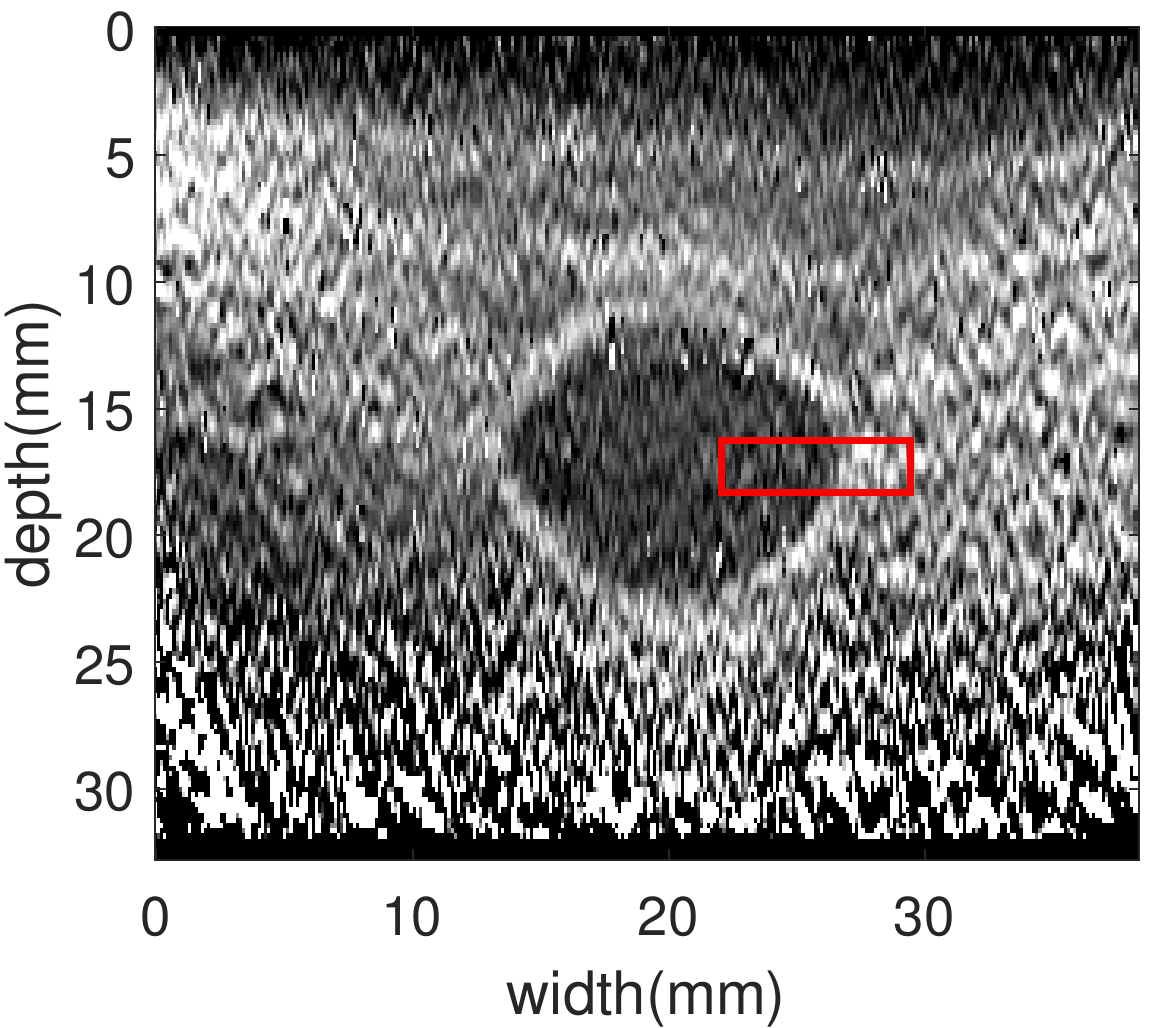} }}%
	\subfloat[]{{\includegraphics[width=3.9cm]{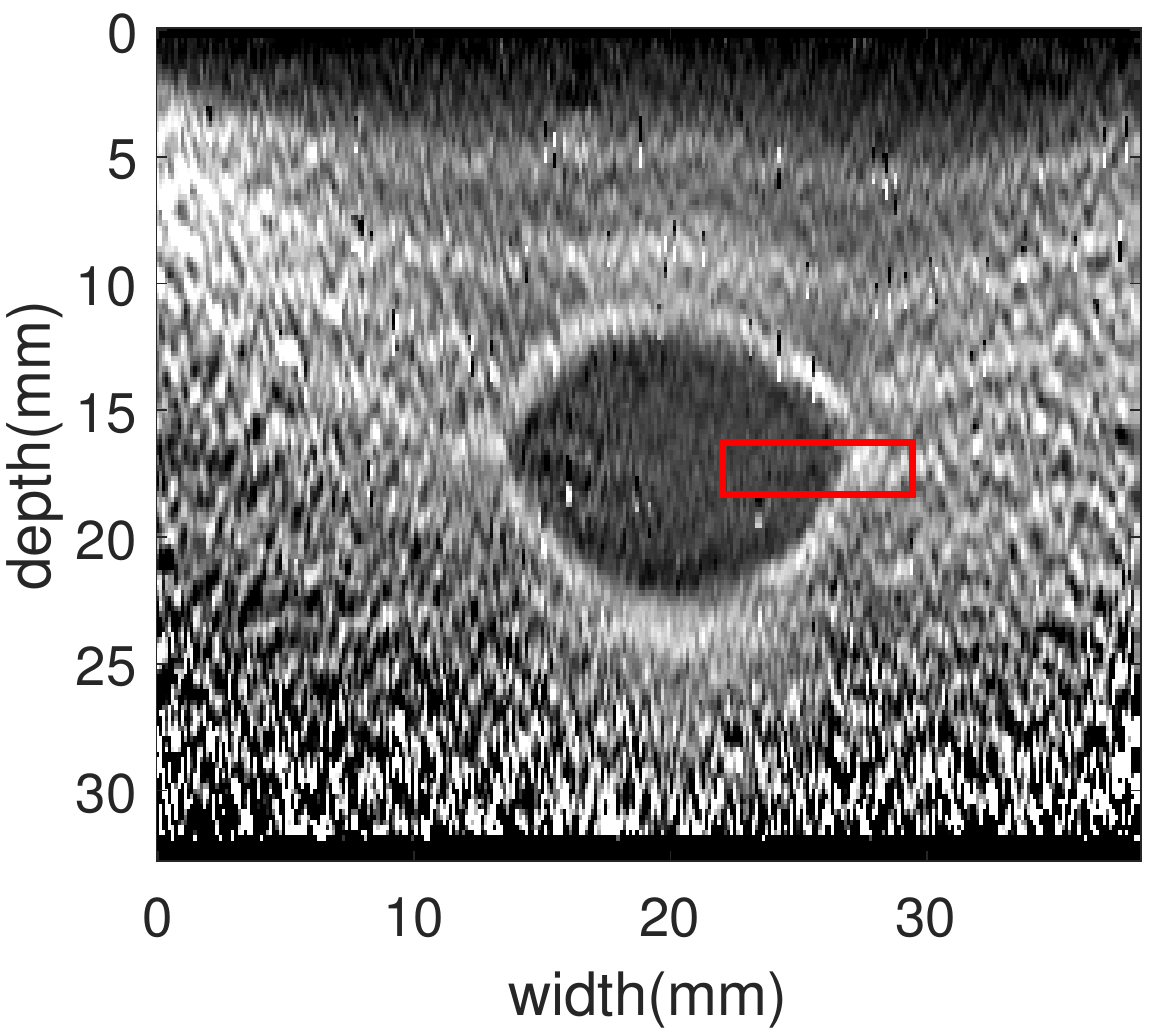} }}
	\qquad
	\qquad
	\subfloat[]{{\includegraphics[width=4.2cm]{phcolo} }}
	\qquad
\qquad
	\subfloat[]{{\includegraphics[width=3.85cm,height=3.5cm]{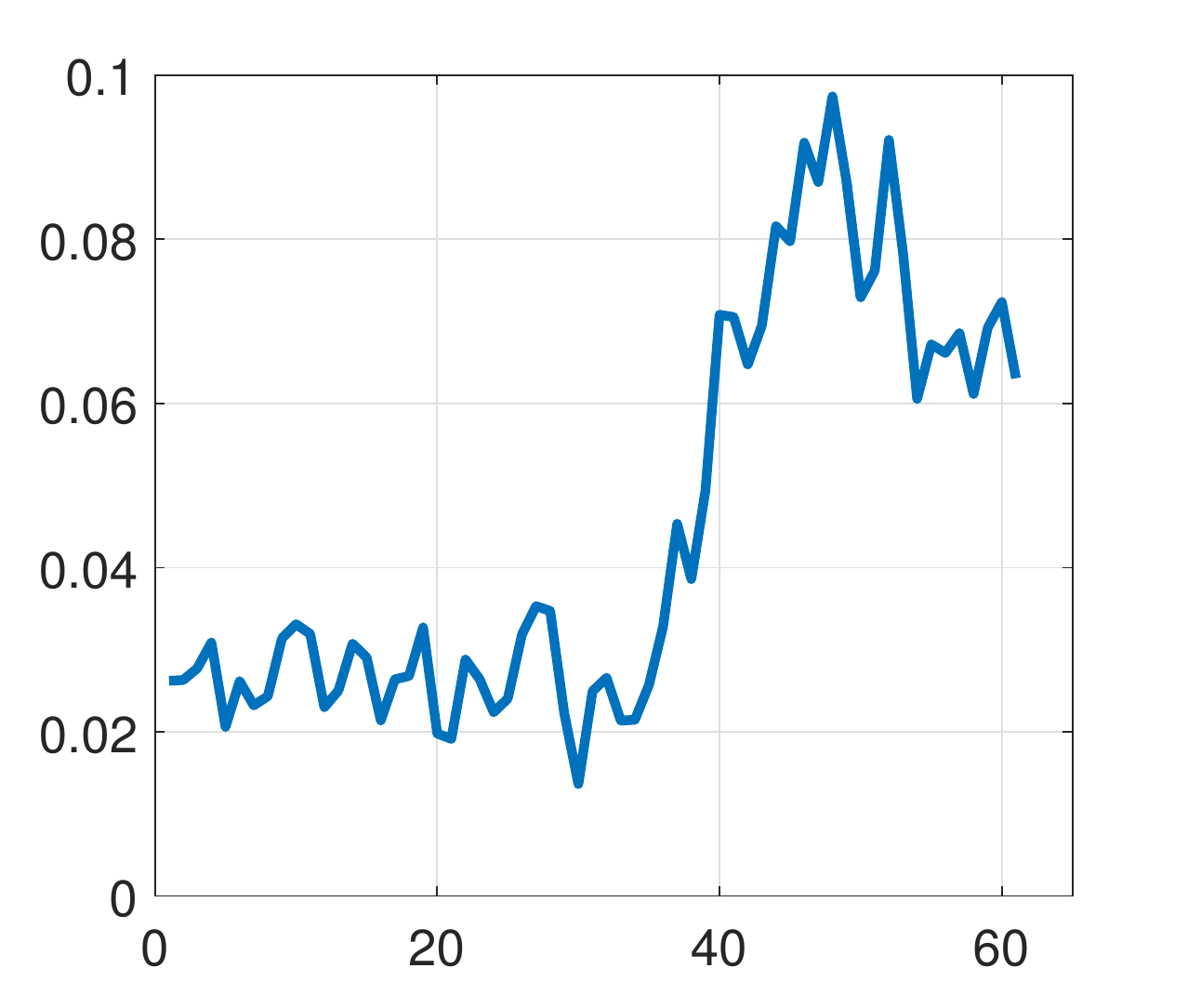} }}%
	\subfloat[]{{\includegraphics[width=3.85cm,height=3.5cm]{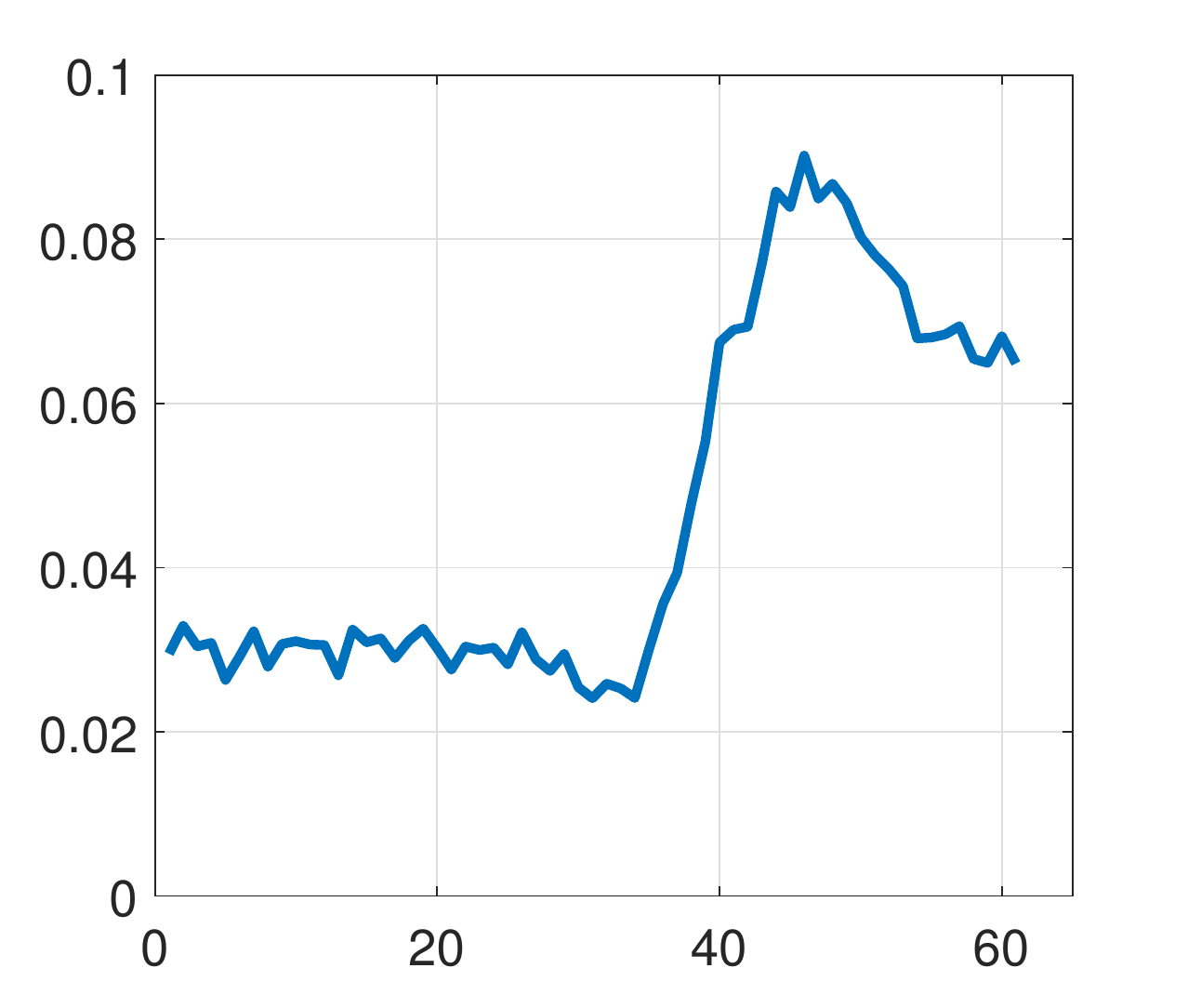} }}
	\caption{Edge spread function for strain images that are calculated using NCC and STNCC. The two red boxes in (a) and (b) show the region of strain image  where edge profiles are  plotted. (d) and (e) show the edge profiles.}
\end{figure}
\subsection*{In-vivo Results}

Two experiments are studied for two different organs of back muscle and liver. 

\subsubsection*{Back Muscle}
\textit{In-vivo} RF data are collected using an ultrasound machine (E-Cube R12, Alpinion, Bothell, WA, USA) with a SC1-4H curvilinear probe at the center frequency of $3.2$ MHz and sampling frequency of $40$ MHz. In this experiment, the probe was hand-held and was placed axially on multifidus muscle while the subject was  lying prone. The subject then performed a contralateral arm lift, which causes deformation (submaximal contraction) in the multifidus muscle. This study was approved by Central Ethics Committee of Health and Social Services from the Ministry of Quebec (MSSS: Ministere de la Sante et des Services Sociaux). The subject provided informed consent for this experiment.

Figure 8 shows B-Mode image of the multifidus muscle, which is delineated by dashed red lines. Figures 9 a-b show the displacement fields estimated with  NCC and STNCC with $70\%$ overlap between windows.  Figure 9 and Table 4 demonstrate that STNCC calculates a superior displacement field compared to NCC.
\begin{figure}
	\centering
	{{\includegraphics[width=7cm]{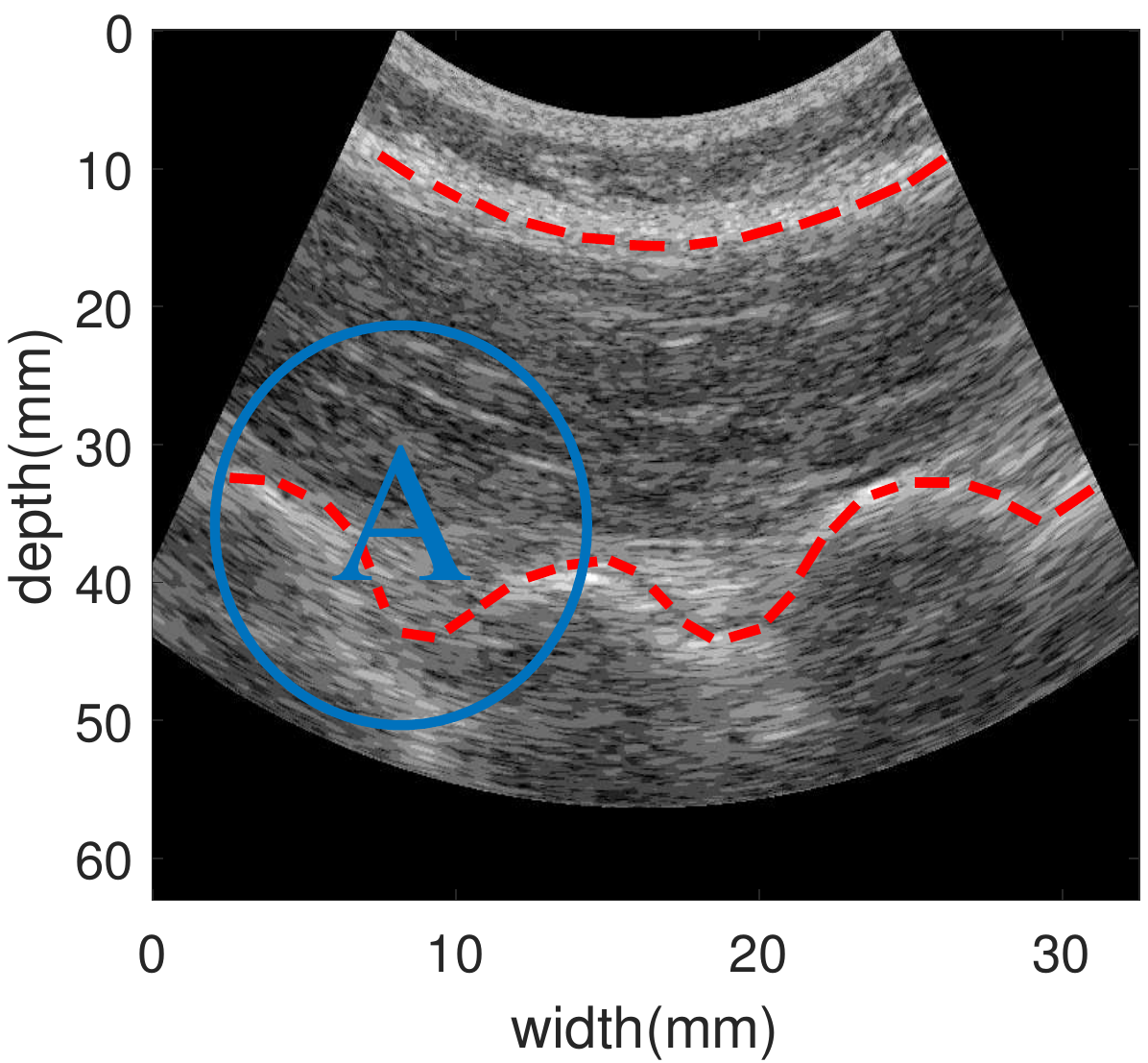} }}%
	\caption{B-mode image of the back muscle. Red dashed lines delineate the multifidus muscle. 
		Visual inspection of the B-mode images shows the maximum displacement occurring in the region marked with the letter A.}	
\end{figure}

\begin{figure}
	\centering
	\subfloat[]{{\includegraphics[width=3.95cm]{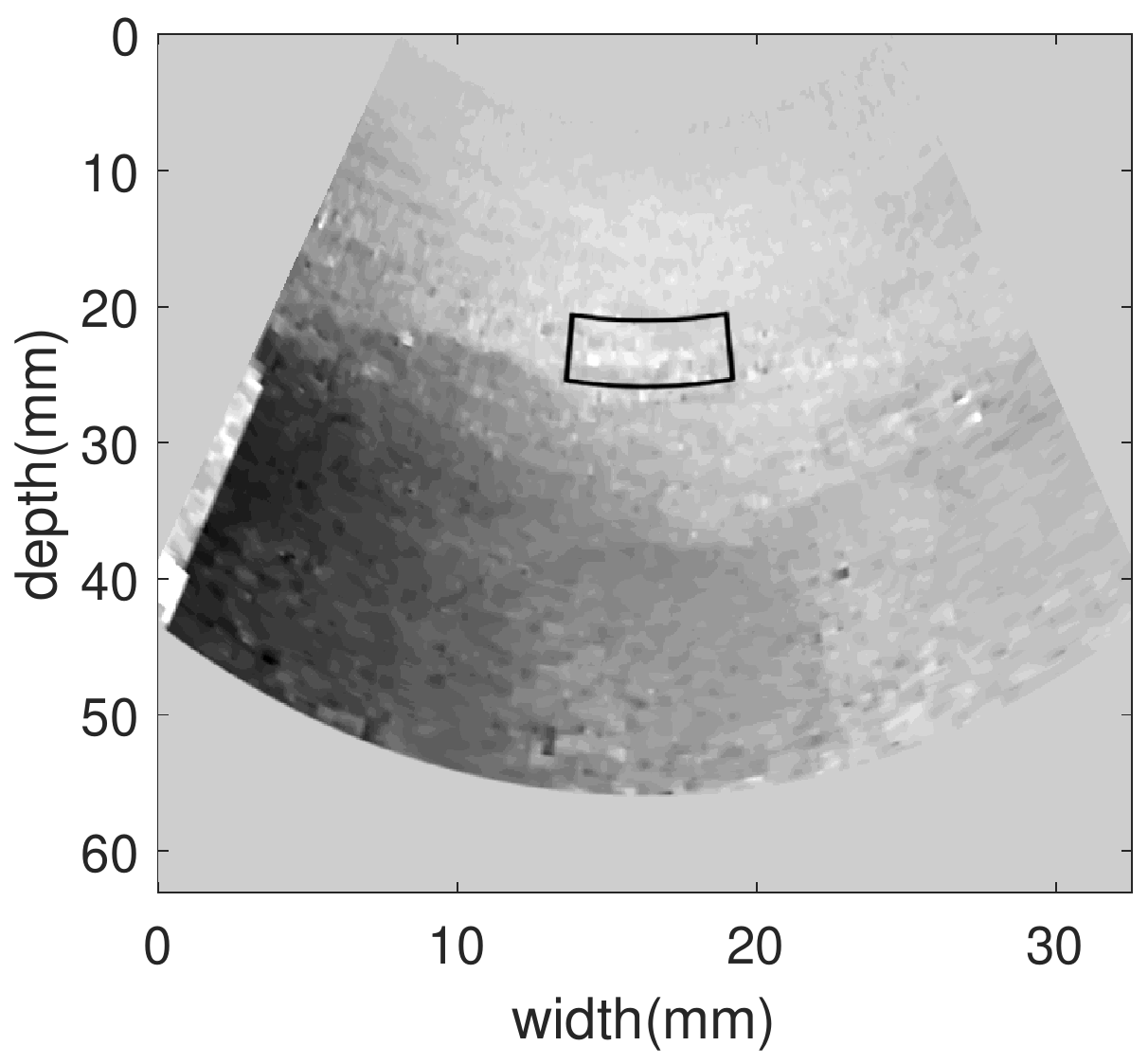} }}%
	\subfloat[]{{\includegraphics[width=3.95cm]{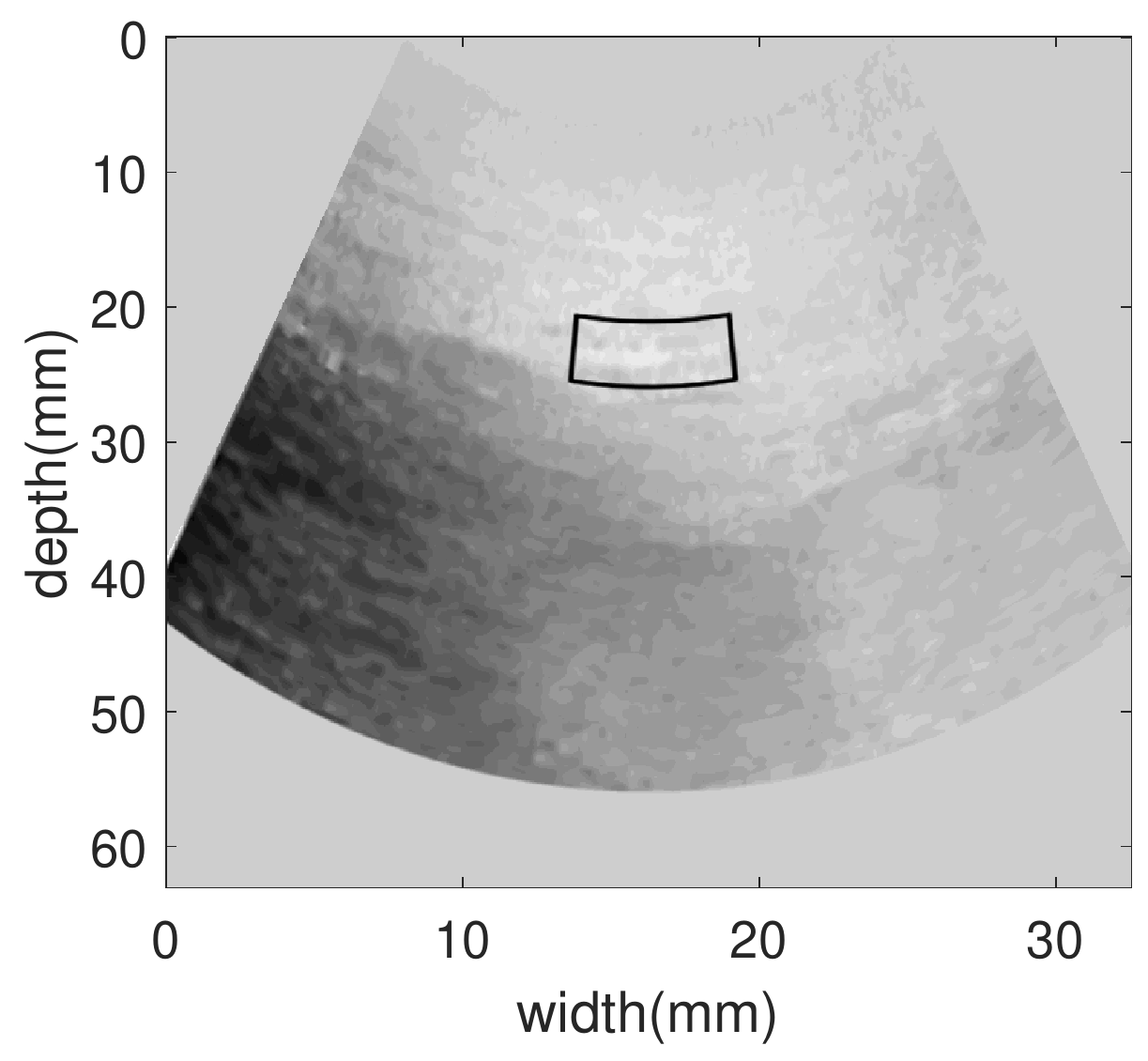} }}
	\qquad
	\subfloat[]{{\includegraphics[width=3.95cm]{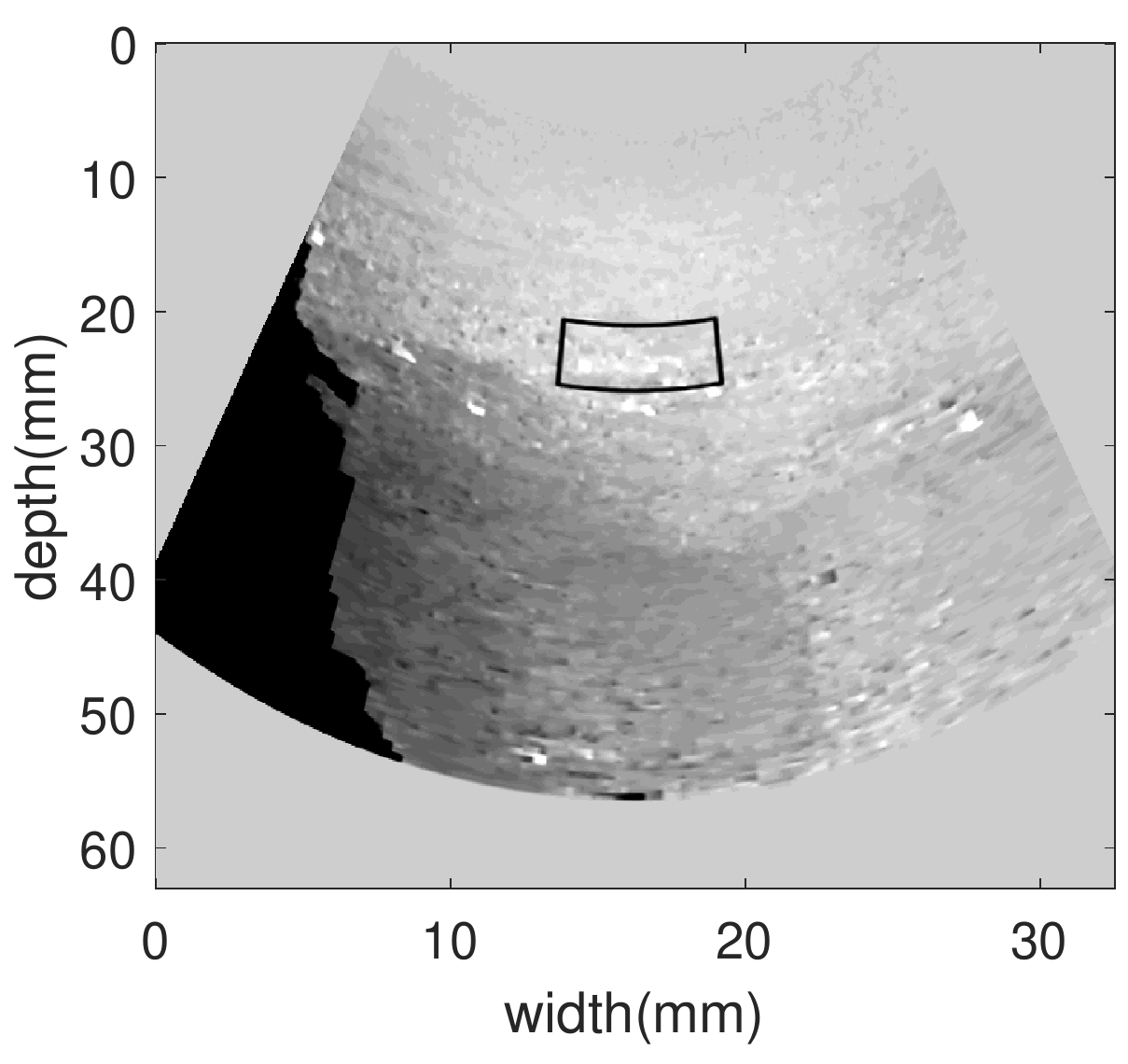} }}%
	\subfloat[]{{\includegraphics[width=3.95cm]{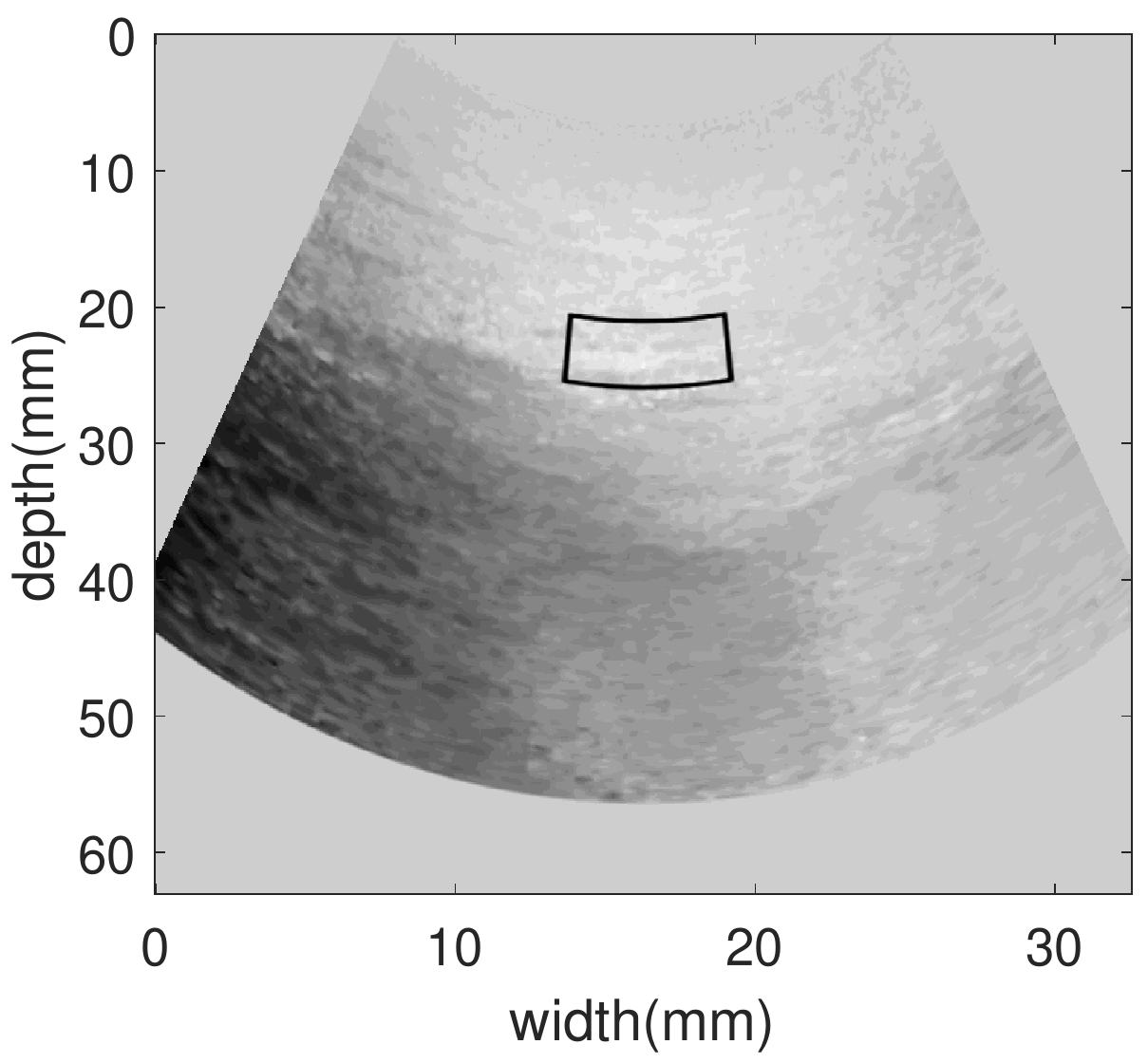} }}
	\qquad
    \qquad
	\subfloat[]{{\includegraphics[width=3.95cm]{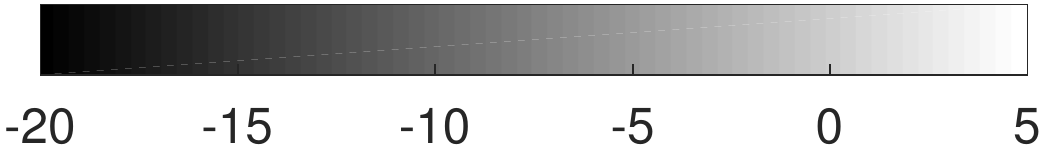} }}
	\caption{Displacement fields of the back muscle  calculated using NCC and
		STNCC. In the first and second rows, the overlap between windows are respectively $70\%$ and $30\%$. The estimated displacement field with NCC is shown in (a) and (c), and the estimated displacement field with STNCC is shown in (b) and (d). 
	}
\end{figure}
We performed another comparison by changing the overlap between consecutive windows.  Figures 9 c-d show  the displacement field estimated with STNCC and NCC with $30\%$ overlap of windows. Comparing Figures 9 a-b and 9 c-d, and also considering Table 4, it is clear that STNCC is substantially less susceptible  to overlap between windows.
\begin{table}
\caption{SNR of displacement images of the  back muscle. The black window is considered in calculating SNR.}
\begin{tabular}[c]{cll}
	\hline		
	Overlap of windows	&	& SNR  \\
	\hline
	\multirow{3}{*}{$\%70$}	&	STNCC    &   1.48    \\
	&	NCC   & 0.60   \\
	&	\textbf{Improvement}   & $\mathbf{\%146.66}$   \\
	\hline
	\multirow{2}{*}{$\%30$}	&	STNCC    &  1.34     \\
	&	NCC   & 0.41   \\
	&	\textbf{Improvement}    & $\mathbf{\%226.82}$   \\
	\hline		
\end{tabular} \\
\end{table}
\subsubsection*{Liver}
The data that  in this experiment is acquired from a patient undergoing open surgical radio frequency thermal ablation for liver cancer before ablation. This data was collected at the Johns Hopkins hospital with an  ultrasound machine (Antares, Siemens, Issaquah, WA, USA) with a VF10-5 linear probe with a center frequency of $6.6$ MHz, sampling frequency of $40$ MHz and frame rate of $30$ fps. The study was approved by the ethics institutional review board at Johns Hopkins. 
Figure 10a shows the B-Mode image, where the tumor is marked with red arrows.
Strain images are computed with NCC and STNCC, and the results are presented in Figure 10b-c.
\begin{figure}
	\centering
	\subfloat[]{{\includegraphics[width=4cm]{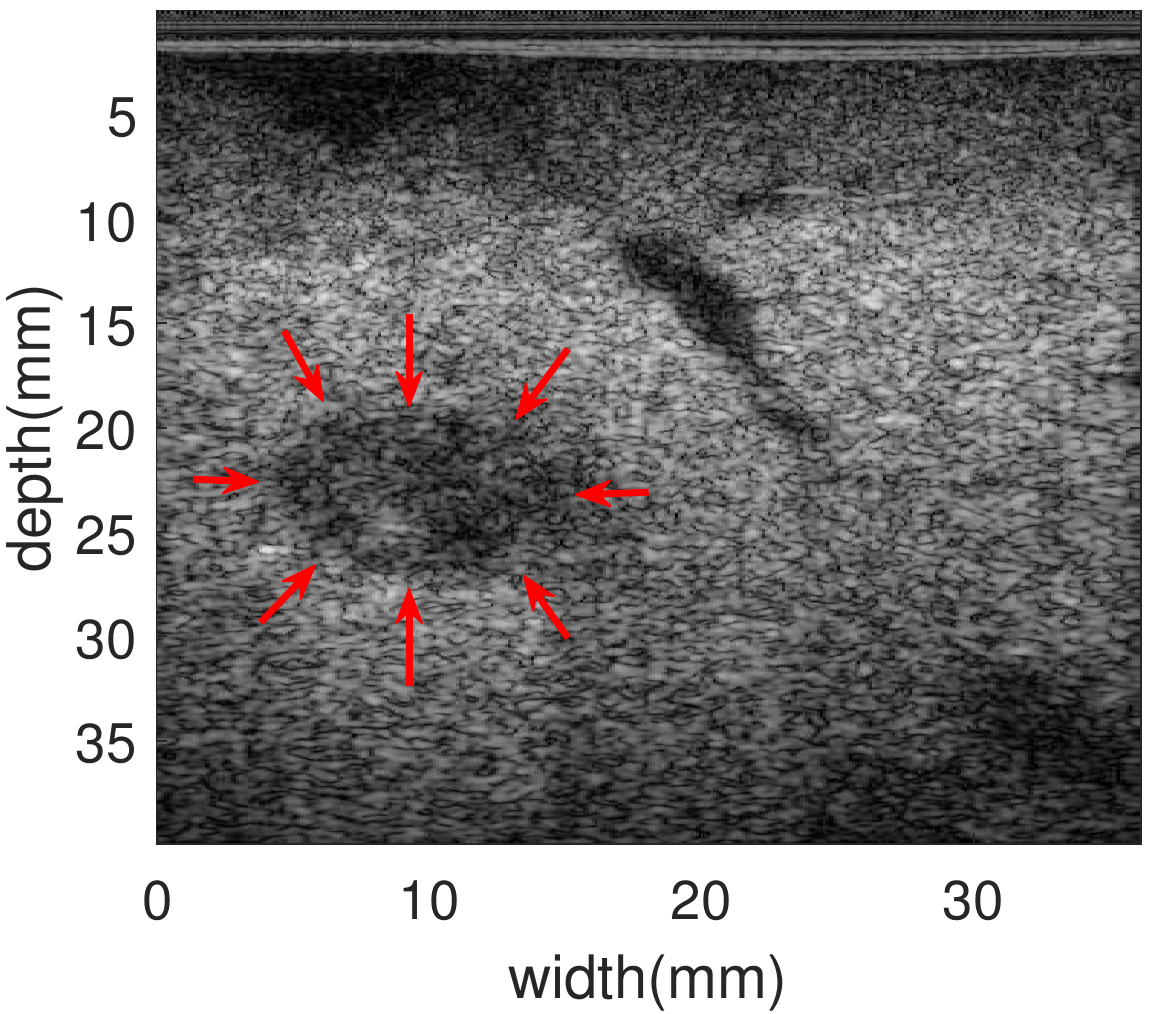} }}%
	\subfloat[]{{\includegraphics[width=3.95cm]{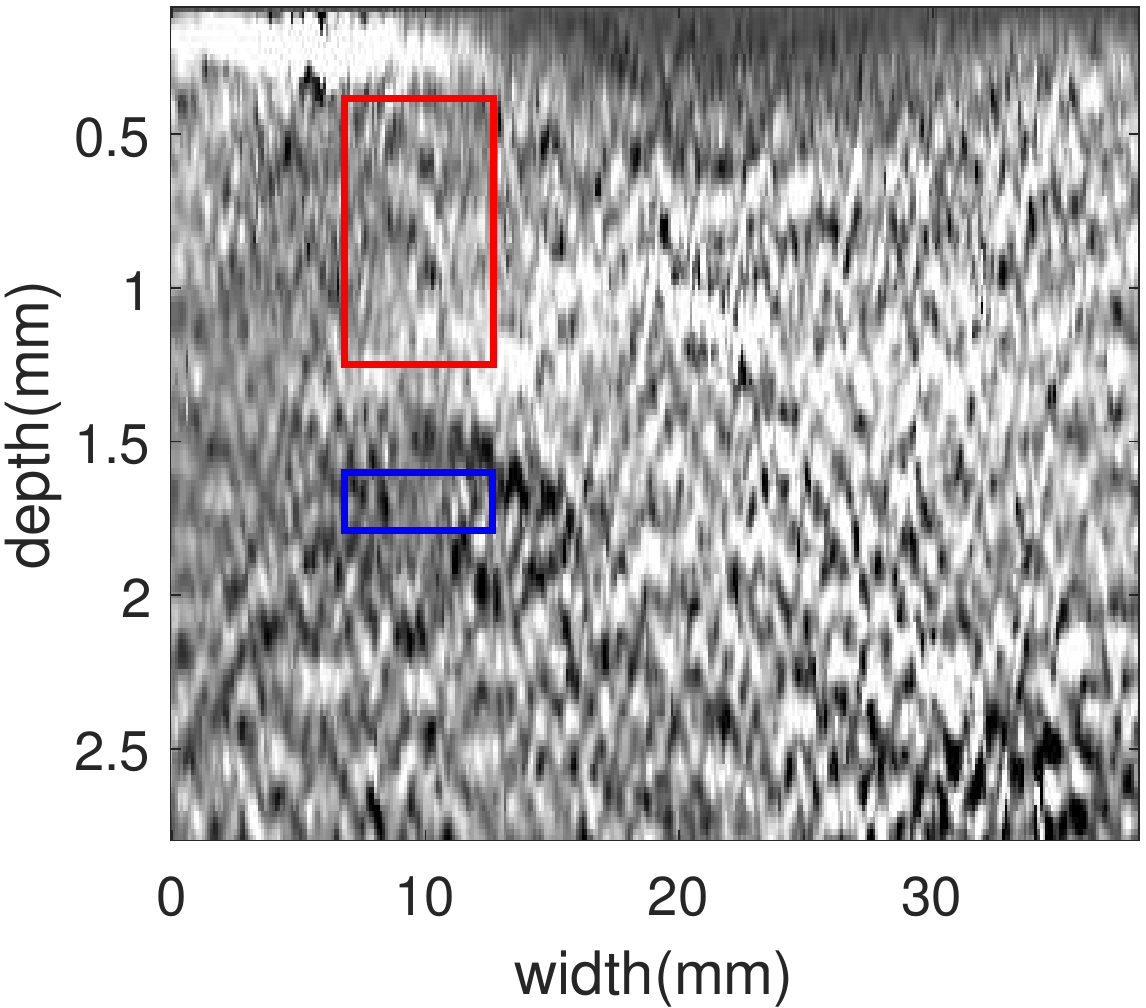} }}%
	\subfloat[]{{\includegraphics[width=3.95cm]{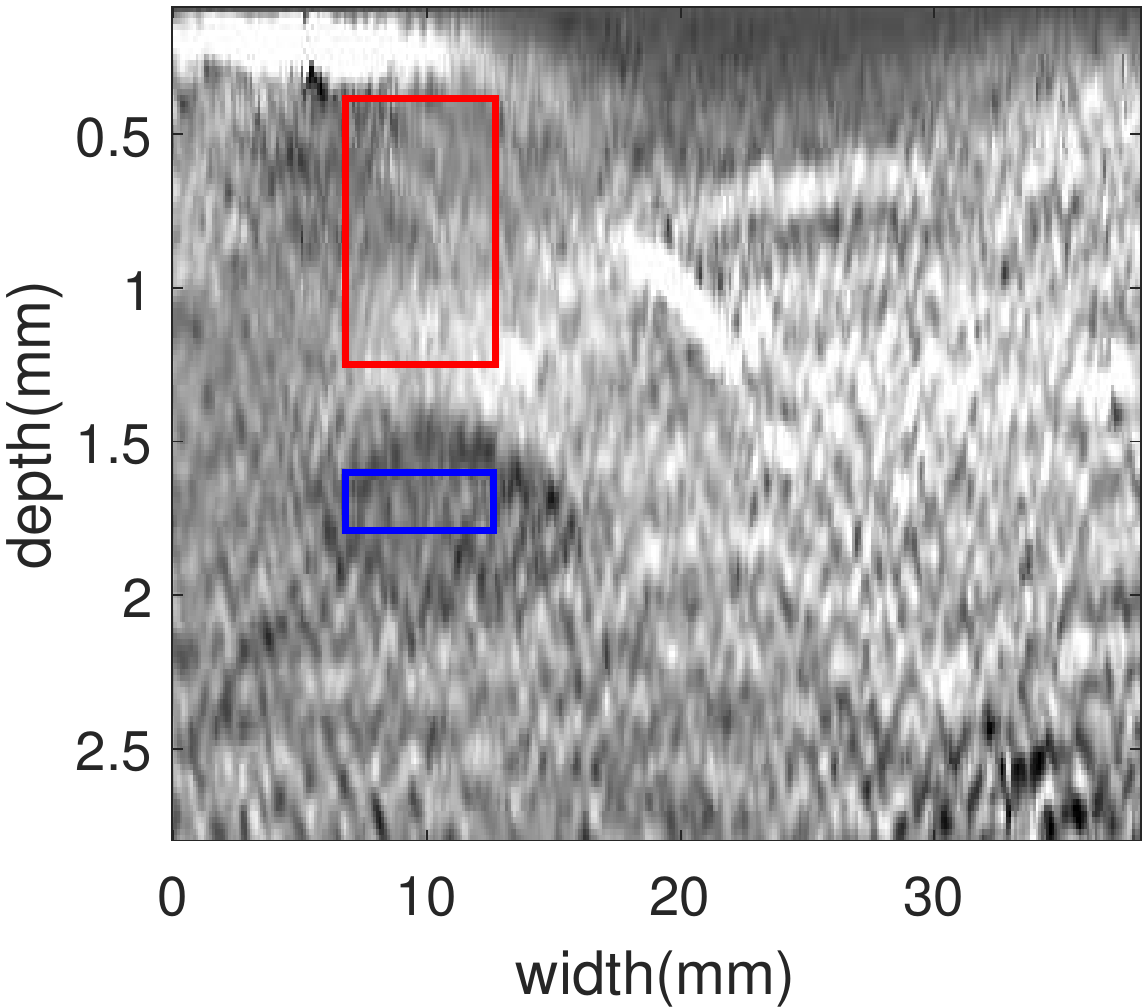} }}
	\qquad
	\subfloat[]{{\includegraphics[width=3.95cm]{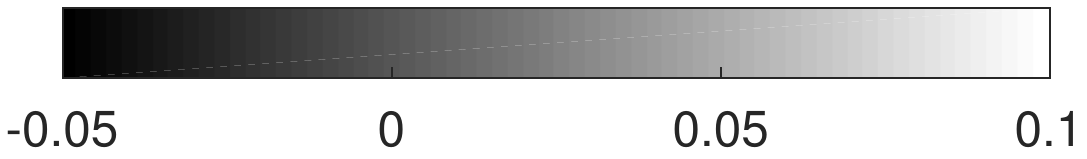} }}
	\caption{B-mode image of the liver with a tumor (marked with red arrows).  Strain images calculated using NCC and
		STNCC are shown in (b) and (c) respectively.
	}
\end{figure}
Visual comparison of the strain images shows that  STNCC generates a strain image with less noise. This is corroborated with quantitative results of Table 5, which shows SNR and CNR. Compared to NCC, STNCSS improves SNR and CNR by respectively $71.06\%$ and $67.15\%$.
\begin{table}
\caption{SNR and CNR  values in  strain images of Figure 11. Windows that are considered for calculating CNR are shown in Figure 11 and only red window is considered for SNR.}
	\begin{tabular}[c]{llr}
		\hline		
		& SNR & CNR \\
		\hline
		STNCC    &   116.77  & 3.46  \\
		NCC   & 68.26  &   2.07   \\
		\textbf{Improvement}   & $\mathbf{\%71.06}$  &   $\mathbf{\%67.15}$   \\
		\hline		
	\end{tabular}\\ 
\end{table}
\section*{Discussion}

Since one sample of RF data is not enough to find displacement map,  window-based techniques assume that the displacement of neighboring samples are the same and look for a similar window in the other image. According to detailed experiments in~\citep{RIGHETTI2002101,lastdis}, assuming that $\lambda$ is one wavelength of ultrasound signal, $10\lambda$  is approximately largest window size for which this assumption is valid. The underlying idea of this project was extending the assumption of spatial continuity to temporal continuity. This is a fair assumption given the high frame rate of ultrasound machines. 

STNCC is more robust to signal de-correlation and can tolerate higher levels of noise compared to NCC. A reason for this improvement is that noise affects different frames by different levels, and by considering multiple frames instead of one, the samples that are less noisy can compensate the effect of noisy samples.

Another advantage of this idea pertains to a wealth of previous work on improving displacement estimation techniques with window-based methods. Future work can focus on applying those methods to 3D windows to further improve the performance of elastography methods. 
Future work can also focus on extracting the  best number of frames to achieve optimal results. 
\section*{Conclusions}
\label{Conclusions}
Ultrasound systems are capable of acquiring images at a very high frame rate. This capability is not exploited in previous window-based elastogrphy algorithms where the windows were only in the spatial domain.
 In this paper, a novel idea was proposed to consider two sequence of images instead of just two images.  In this method, spatio-temporal windows in the first series of images are matched to those of the second series of images. It was shown using simulation, phantom and \textit{in-vivo} experiments  that extension of windows in the temporal direction substantially improves the quality of displacement estimation.

\section*{Acknowledgments}
\label{Ack}
This work was supported by Natural Science and Engineering
Research Council of Canada (NSERC) Discovery Grants RGPIN-2015-04136 and RGPIN-2017-06629.
The  \textit{in-vivo} data of liver patient was collected at Johns
Hopkins Hospital. Authors would like to thank the principal investigators
Drs. E. Boctor, M. Choti and G. Hager for sharing the data with us. The RF data of back muscle was collected at Concordia University's PERFORM Centre with an Alpinion ultrasound machine. The authors would like to thank Julian Lee from Alpinion USA for his technical help.





\pagebreak

\bibliographystyle{UMB-elsarticle-harvbib}
\bibliography{UMB-elsarticle-template-harv_HR_April3}




\pagebreak





\pagebreak



\pagebreak



\end{document}